\documentclass[review,10pt]{elsarticle}
\usepackage{parskip}
\usepackage{newtxmath}
\usepackage{graphicx}
\usepackage{hyperref}
\usepackage{bm}
\usepackage{amsmath}
\usepackage[version=4]{mhchem}
\usepackage{color}
\usepackage{upgreek}
\usepackage{wrapfig}
\usepackage[export]{adjustbox}
\usepackage{caption}
\usepackage{enumitem}
\usepackage{float}
\usepackage{tikz}
\usetikzlibrary{positioning}
\usetikzlibrary{plotmarks}
\usetikzlibrary{trees, arrows.meta, decorations.pathmorphing}
\usepackage{pgfplots} 
\usepackage{pdflscape}
\pgfplotsset{
    try min ticks=6,
        layers/3_layers/.define layer set={
            background,
            main,
            foreground
        }{
        },
        set layers=3_layers,
    grid style={gray,opacity=0.08},
    major grid style={gray,line width=1,opacity=0.2},
    every axis plot post/.append style={every mark/.append style={mark size=1.9pt,line width=0.05pt,draw opacity=0.7}},
    legend style={font=\small}
}
\pgfplotsset{compat=newest} 
\pgfplotsset{plot coordinates/math parser=false}

\newlength\fheight
\newlength\fwidth 

\usepackage{tabularx}
\usepackage{longtable}
\usepackage{booktabs}
\usepackage{mathtools}
\usepackage{setspace}
\AtBeginDocument{%
	\setlength\abovedisplayskip{10pt}
	\setlength\belowdisplayskip{10pt}}

\allowdisplaybreaks
\usepackage[a4paper,
            bindingoffset=0.1in,
            left=0.55in,
            right=0.65in,
            top=0.53in,
            bottom=0.56in,
            footskip=.3in]{geometry}
\usepackage{layouts}

\biboptions{sort&compress}
\journal{Results in Engineering}

\usepackage{setspace}

\usepackage{placeins}

\let\Oldsection\section
\renewcommand{\section}{\FloatBarrier\Oldsection}

\let\Oldsubsection\subsection
\renewcommand{\subsection}{\FloatBarrier\Oldsubsection}

\let\Oldsubsubsection\subsubsection
\renewcommand{\subsubsection}{\FloatBarrier\Oldsubsubsection}

\usepackage{xspace}

\newcommand{\Fetwp}{\ce{Fe^{2+}}\xspace}
\newcommand{\Fethp}{\ce{Fe^{3+}}\xspace}
\newcommand{\Fetwpc}{\ce{Fe^{II}}\xspace}
\newcommand{\Fethpc}{\ce{Fe^{III}}\xspace}
\newcommand{\FetOt}{\ce{Fe2O3}\xspace}
\newcommand{\FetOf}{\ce{Fe3O4}\xspace}
\newcommand{\FeO}{\ce{FeO}\xspace}
\newcommand{\Fe}{\ce{Fe}\xspace}
\newcommand{\aFe}{\ce{\alpha-Fe}\xspace}

\newcommand{\aFetOt}{\ce{\alpha-Fe2O3}\xspace}
\newcommand{\gFetOt}{\ce{\gamma-Fe2O3}\xspace}

\newcommand{\POnepf}{\ce{\Phi=1.5}\xspace}

\newcommand{\specialcellc}[2][c]{%
  \begin{tabular}[#1]{@{}c@{}}#2\end{tabular}} %

\begin{document}
\begin{frontmatter}
\title{Influence of light, temperature, and iron oxidation state on the dissolution rate of combusted iron particles in oxalic acid}
\author[add_a]{M. Lausch\texorpdfstring{\corref{cor1}}{}}
\ead{lausch@sla.tu-darmstadt.de}
\author[add_a]{Y. Ruan}
\author[add_a]{P. Brockmann}
\author[add_d,add_e]{A. Zimina}
\author[add_b,add_c]{B.J.M. Etzold}
\author[add_a]{J. Hussong}
\cortext[cor1]{Corresponding author}
\address[add_a]{Technische Universit\"{a}t Darmstadt, Institute for Fluid Mechanics and Aerodynamics; 64347 Griesheim}
\address[add_b]{Technische Universität Darmstadt, Ernst-Berl-Institute for Technical Chemistry and Macromolecular Science; 64287 Darmstadt, Germany.}
\address[add_c]{Friedrich-Alexander-Universität Erlangen-Nürnberg, Power-To-X Technologies; 90762 Fürth, Germany.}
\address[add_d]{Karlsruhe Institute for Technology, Institute of Catalysis Research and Technology; 76344 Eggenstein-Leopoldshafen, Germany }
\address[add_e]{Karlsruhe Institute for Technology, Institute for Chemical Technology and Polymer Chemistry; 76131 Karlsruhe, Germany}
\address[add_f]{Technische Universit\"{a}t Darmstadt, Institute for Catalysts and Electrocatalysts; 64287 Darmstadt, Germany.}

\begin{abstract}
It is essential to control the dissolution rate of iron oxide particles for a prospective acidic iron electrowinning process. In this study, the combined influence of temperature (40–80$^\circ$C) and short-wavelength light exposure on the dissolution rate of combusted iron particles in aqueous oxalic acid (0.45 mol/L) is experimentally investigated. The combusted iron particles were produced with various fuel-to-air equivalence ratios during combustion. Unlike previous dissolution studies on single\textendash phase iron oxides, these particles comprise a heterogeneous mixture of iron oxides\,\textendash\,primarily hematite and magnetite. In situ video recordings revealed the evolution of the particle size and morphology during dissolution. Increasing the temperature accelerated the reaction rate, and an additional light\textendash induced enhancement became significant only above 40$^\circ$C for the duration of the experiments. This behavior differs significantly from that observed for hematite/maghemite mixed oxides and is attributed to the internal hematite and magnetite structure of the combusted iron particles. At 80$^\circ$C under short-wavelength light irradiation, a sudden decrease in the reaction rate was observed owing to solid ferrous oxide formation. Although the fuel-to-air ratio affected the iron oxide composition inside the particles, it did not significantly affect the dissolution rate of the combusted iron particles.
\end{abstract}


\end{frontmatter}
\section{Introduction}
\label{sec:Introduction}
Over 80\% of the energy required for steel production is derived from fossil fuels, with 75\% being coal\textendash based~\citep{IEA2022}. Consequently, the steel industry is a significant contributor to \ce{CO2} emissions, which will continue to be a concern for the foreseeable future. Processes such as electrowinning (EW), smelting~\citep{MEIJER2013}, and electric arc furnaces~\citep{HASANBEIGI2014,HARVEY2021} could significantly lower future carbon emissions, especially considering the scale of steel production (reaching 1.9 Gt in 2023~\citep{WORLDSTEEL2023}). For future iron EW, slurry electrolysis is the most widely pursued method (see~\ref{sec:Iron_electrowinning}). In this process, the electrolyte also serves as a leaching agent \citep{LIU2024,SIWAL2023} by dissolving suspended iron oxides while the resulting iron ions are simultaneously deposited at the cathode. Electrowinning also aligns with the concept of an iron energy cycle, where iron dust combustion releases energy, and the resulting iron oxide particles are later reduced again to iron~\citep{DEBIAGI2022,BERGTHORSON2015,BERGTHORSON2018,TOTH2020,DIRVEN2018}. In such a system, the metal acts as a chemical energy carrier for renewable energy, a process also denoted as Metal\textendash enabled Cycle of Renewable Energy (MeCRE). During the combustion process, the particle size, temperature, and oxygen availability determine whether heat and mass transfer in the particle boundary layer, intra\textendash particle processes, or condensed phase equilibrium conditions govern the oxidation rate of the particles~\citep{MI2022,MICH2024}. Most combusted iron particles (CIPs) are spherical, some containing internal pores~\citep{Li2021,CHOISEZ2022,POLETAEV2022,DEUTSCHMANN2024} that comprise mostly hematite (\aFetOt) and magnetite (\FetOf) with traces of wüstite (\FeO) and uncombusted iron~\citep{BUCHHEISER2023,CHOISEZ2022}. The focus of this study is on the dissolution of these CIPs as part of the electrochemical reduction process. 

Increasing the dissolution rate is desirable for an acidic EW process because it allows for higher limiting currents and deposition rates. Experiments were performed in aqueous oxalic acid (OxA) because it provides one of the highest tested dissolution rates~\citep{AMBIKADEVI2000,BORGHI1996,TAXIARCHOU1997b} along with a high reducing power~\citep{LEE2007} compared to other organic acids and high iron oxide solubility (OxA/iron molar ratio of approximately 1.82 \citep{SANTAWAJA2021}). The dissolution of the iron oxides in OxA can be controlled by tuning temperature and the presence of \Fetwp ions in the solution. The latter can be influenced by light irradiation due to photoactive complexes in solution. Unlike pure iron oxides commonly investigated in the literature, CIPs exhibit distinct macroscopic and microscopic features. Therefore, the combined effects of thermal and radiative conditions and CIPs properties on the dissolution rate and intermediate reactions are systematically analyzed. Developing a comprehensive understanding of the dissolution mechanisms within aqueous systems may enable the transfer of these insights to non\textendash aqueous processes. OxA can also be part of non\textendash aqueous electrolytes, such as deep eutectic solvents~\citep{SMITH2014,ABBOTT2004}, which suppress the hydrogen evolution and yield high\textendash quality metal deposits.  Although iron experiments remain to be conducted, choline chloride:OxA mixtures have shown promising results for the EW of bismuth, tellurium~\citep{AGAPESCU2013,SORGHO2022}, and gold~\citep{SOMA2024}.

Generally, the rate at which a particle dissolves is controlled either by solute transport (within the particle or across its concentration boundary layer) or the surface reaction~\citep{RUDZINSKI2007,PLAZINSKI2009}. For the dissolution of iron oxides in OxA, the overall dissolution process is usually modeled to be controlled by surface reactions~\citep{SUTER1988,BLESA1987,LITTER1991,STUMM1992} and thus depends only on the reaction kinetics and available surface area. Consequently, the dissolution rate per unit surface area is independent of the particle size, and hydrodynamic conditions do not directly affect this rate. However, they can indirectly influence dissolution by promoting particle break\textendash up and altering available surface area~\citep{ZHAO2018,LAUSCH2024}. The assumption of surface\textendash reaction controlled dissolution may become invalid for particles at the nanometer scale, where the anisotropy of the terminally coordinated \Fe ions and the connection between single crystal faces~\citep{PRACHI2016,YANINA2008} due to the semiconducting properties of iron oxides becomes more important.\\
Many studies detail the interactions between OxA and iron oxides and classify the related reaction mechanisms~\citep{PANIAS1996A,STUMM1992,LAUSCH2024,LV2022}. Fig.~\ref{fig:schematic_mechanisms} shows an overview of the occurring reactions and influencing factors that are known from the literature. 
The dissolution of iron oxides in OxA can proceed via two pathways~\citep{LITTER1991,PANIAS1996A}, which are distinguished by the reduction of iron in the crystal before dissolution. Numerous factors influence the contribution of the two pathways to the overall dissolution reaction through related reaction mechanisms and the formation of iron\textendash oxalate complexes, including temperature, pH, light exposure, and the concentration of ferrous ions in the solution~\citep{LV2022,PANIAS1996A}. The associated equations and references are provided in \ref{sec:appendix_mechanisms_equations}. For the dissolution of CIPs specifically, the formation of solid iron oxalate dihydrate $\big(\mathrm{Fe}^{\mathrm{II}} \mathrm{C}_2 \mathrm{O}_4 \cdot 2\,\mathrm{H}_2 \mathrm{O}\big)$ is relevant, because it is an undesirable byproduct that binds some of the iron and needs to be converted to \Fe e.g., through pyrolytic reduction~\citep{SANTAWAJA2020}. The solubility limit of this ferrous oxalate in water at room temperature is $c_{\mathrm{FeC_2O_4}}\approx 447.2\,\upmu$M~\citep[p.487]{HOGNES1940}. 

\begin{figure*}[!ht]
    \centering
    \begin{tikzpicture}[
    light/.style={rectangle,draw,fill=white,align=center,rounded corners=.8ex,fill opacity=1,text opacity=1},
    nolight/.style={fill=white,align=center,fill opacity=1,text opacity=1},
    first/.style={sibling distance=7.0cm,level distance=0.5cm},
    second/.style={sibling distance=2.9cm,level distance=1.4cm,
    edge from parent fork down,
    edge from parent path={(\tikzparentnode.south) -- ++(0,-0.35cm) -| (\tikzchildnode)}}]
    \footnotesize

    \newcommand{\largerectx}{0}
    \newcommand{\largerecty}{0}
    \newcommand{\largerectwidth}{16}
    \newcommand{\largerectheight}{6.4}
    
    \newcommand{\smallerrectx}{0.5}
    \newcommand{\smallerrecty}{1.25}
    \newcommand{\smallerrectwidth}{15}
    \newcommand{\smallerrectheight}{3.9}

    \draw[rounded corners, thick] (\largerectx,\largerecty) rectangle 
    ++(\largerectwidth,\largerectheight);

    \node at (\largerectx + \largerectwidth/2, \largerecty + \largerectheight + 0.52) {Increasing pH, added \Fetwp in solution};
    
    \draw[-{Stealth[length=3mm, width=2mm]}, thick] (\largerectx + \largerectwidth/2 - 4, \largerecty + \largerectheight + 0.25) -- ++(8, 0); 
    
    \node at (\largerectx + 1*\largerectwidth/8 - 1.1, \largerecty + \largerectheight - 0.3) 
    {\textbf{Liquid}}; 

    \draw[rounded corners, dashed, thick] (\smallerrectx,\smallerrecty) rectangle 
    ++(\smallerrectwidth,\smallerrectheight);
    
    \node at (\smallerrectx + 1*\smallerrectwidth/8 - 1.2, \smallerrecty + \smallerrectheight - 0.3) 
    {\textbf{Solid}}; 

    \node[align=right] at (\largerectx + 1*\largerectwidth/2 - 0.3, \largerecty + \largerectheight - 0.45) {$^{\mathrm{\textbf{a}}}$OxA dissociation, Eq.~\ref{equ:chem_acid_dissociation}\\ion species distribution };

    \node[align=right] at (\largerectx + 4*\largerectwidth/5+1.3, \largerecty + \largerectheight - 0.45) {Precipitation, Eq.~\ref{equ:chem_iron_oxalate_synthesis}\\$\big(\mathrm{Fe}^{\mathrm{II}} \mathrm{C}_2 \mathrm{O}_4 \cdot 2\,\mathrm{H}_2 \mathrm{O}\big)$};
    
    \draw[-{Stealth[length=3mm, width=2mm]}, thick]  (\largerectx + 4*\largerectwidth/5 + 1.3, \largerecty + \largerectheight - 0.8) -- ++(0, -0.85);
    
    \node[align=center, fill=white,fill opacity=1,text opacity=1] at (\smallerrectx + 1*\smallerrectwidth/2 - 0.4, \smallerrecty + \smallerrectheight) 
    {Additional surface creation through particle breakup~\citep{LAUSCH2024}}; 

    \node[draw,align=left,rounded corners, dashed, thick] at (\smallerrectx + 1*\smallerrectwidth/10 - 0.33, \smallerrecty + 3.3*\smallerrectheight/5 + 0.03) {Rhombohedral\\crystal structure};

    \node[draw,align=right,rounded corners, dashed, thick] at (\smallerrectx + 9*\smallerrectwidth/10 + 0.33, \smallerrecty + 3.3*\smallerrectheight/5 +0.03) {Spinel\\crystal structure};

    \node[align=center,draw,rounded corners=.8ex] at (\largerectx + 2*\largerectwidth/3 + 0.3, \largerecty + 4.4 - 0.2) {internal charge transfer\\(semiconductor~\citep{PRACHI2016,YANINA2008,SHERMAN2005})};

    \draw[decorate, decoration={snake, amplitude=0.3mm, segment length=1mm, post length=1mm}, -{Stealth[scale=0.8]}] (\largerectx + 2*\largerectwidth/3 + 0.4 +       2.3, \largerecty + 4.6 -0.3 +     0.52) -- ++(-0.4, -0.4 *1.3/2.0); 

    \node[align=right] at (\largerectx + 2*\largerectwidth/3 +  2.3, \largerecty + 4.6 -0.3 +     0.49) {\tiny UV};

    \begin{scope}[shift={(\smallerrectx + \smallerrectwidth/2 -0.4, \smallerrecty + 1*\smallerrectheight/4 + 1.6)}]
    \coordinate
    node[align=center] {\textbf{Protonation}\\Eq.~\ref{equ:chem_surface_protonisation}}
    [edge from parent fork down, edge from parent path={(\tikzparentnode.south) -- ++(0,-0.3cm) -| (\tikzchildnode.south)}]
    child[first] {node {Non-reductive}
      child[second] {node[nolight] {\textbf{Direct detachment}\\see \citep{STUMM1987}\\cited in \citep[p.300]{CORNELL2003}}}
      child[second] {node[nolight] {$^{\mathrm{\textbf{a}}}$\textbf{Complexation}\\Eq.~\ref{equ:chem_nonreductive_dissolution}}}
      }
    child[first]{node {Reductive}
      child[second] {node[nolight] {$^{\mathrm{\textbf{a}}}$\textbf{Complexation}\\Eq.~\ref{equ:chem_reductive_dissolution_induction}}}
      child[second] {node[light]   {\textbf{Photochemical}\\ \scriptsize direct~\citep{SHERMAN2005}\\\scriptsize ligand-promoted~\citep{STUMM1992,LV2022}}}
      child[second] {node[nolight] {$^{\mathrm{\textbf{b}}}$\textbf{Autocatalytic}\\Eq.~\ref{equ:chem_reductive_dissolution_autocatalytic}}}
      };
    \end{scope}

    \draw[decorate, decoration={snake, amplitude=0.3mm, segment length=1mm, post length=1mm}, -{Stealth[scale=0.8]}] (\largerectx + 2*\largerectwidth/3 + 2.37, \largerecty + 1.8 + 0.9) -- ++(-0.4, -0.4 *1.3/2.0); 
    \node[align=right] at (\largerectx + 2*\largerectwidth/3 + 1.97, \largerecty + 1.8 + 0.84) {\tiny UV};

    \draw[-{Stealth[length=3mm, width=2mm]}, thick]  (\largerectx + 2.65, \smallerrecty + 0.1) -- ++(0, -0.5);

    \draw[-{Stealth[length=3mm, width=2mm]}, thick]  (\largerectx + 5.55, \smallerrecty + 0.35) -- ++(0, -0.75);

    \draw[-{Stealth[length=3mm, width=2mm]}, thick]  (\largerectx + 8.2, \smallerrecty + 0.35) -- ++(0, -0.75);

    \draw[-{Stealth[length=3mm, width=2mm]}, thick]  (\largerectx + 11.1, \smallerrecty + 0.1) -- ++(0, -0.5);

    \draw[-{Stealth[length=3mm, width=2mm]}, thick]  (\largerectx + 13.9, \smallerrecty + 0.3) -- ++(0, -0.7);

    \draw[-{Stealth[length=3mm, width=2mm]}, thick]  (\largerectx + 14.2, \smallerrecty - 0.4) -- ++(0, 0.7);

    \node[align=center] at (\largerectx + 1*\largerectwidth/2 + 0.1, \largerecty + 0.93*\largerectheight/15) {$^{\mathrm{\textbf{b}}}$Stable Fe-Ox complex\\species distribution~\citep{PANIAS1996A,LAUSCH2024}, Fig.~\ref{fig:speciation_actual_concentration},~\ref{fig:speciation_all_fe2_fe3}};

    \node[draw,rounded corners=.8ex] at (\largerectx + 7*\largerectwidth/8 - 0.55, \largerecty + 0.93*\largerectheight/15) {$^{\mathrm{\textbf{b}}}$Complex photolysis, Eq.~\ref{equ:chem_photolysis_iron_oxalate_complex}};
    
    \node[align=left] at (\largerectx + 1*\largerectwidth/8 - 0.15, \largerecty + 0.93*\largerectheight/15) {$^{\mathrm{\textbf{b}}}$Iron oxidation, Eq.~\ref{equ:chem_ferrous_sol_re-oxidation}};
    
    \draw[decorate, decoration={snake, amplitude=0.3mm, segment length=1mm, post length=1mm}, -{Stealth[scale=0.8]}] (\largerectx + 7*\largerectwidth/8 - 0.4 +       2.29, \largerecty + 0.93*\largerectheight/15 +     0.46) -- ++(-0.4, -0.4 *1.3/2.0); 
    \node[align=right] at (\largerectx + 7*\largerectwidth/8 - 0.8 +       2.29, \largerecty + 0.93*\largerectheight/15 +  0.42) {\tiny UV};

    \draw[-{Stealth[length=3mm, width=2mm]}, thick] (\largerectx + \largerectwidth/2 + 4, \largerecty - 0.25) -- ++(-8, 0); 

    \node at (\largerectx + \largerectwidth/2, \largerecty - 0.52) {Increasing temperature, OxA concentration};
    
\end{tikzpicture}
\caption{Schematic overview of the occurring reaction mechanisms and influencing factors. The lateral position indicates the associated dominant process (left - non-reductive; right - reductive). After protonation, all subsequent mechanisms can operate concurrently. The oxidation number of a lattice ion is represented using Roman numerals, while Arabic numerals are used for all other forms.}
\label{fig:schematic_mechanisms}
\end{figure*}

Given their significance for this study, a detailed summary of the influence of temperature and light on the dissolution process is provided. Many studies consistently demonstrate that an increase in temperature accelerates dissolution regardless of the iron oxide used~\citep{VEHMAANPERAE2021A,MARTINEZLUEVANOS2011,SALMIMIES2016,SALMIMIES2012,LEE2006}. Owing to the activation energy required for the surface complex to detach from the oxide \citet{PANIAS1996A} propose that non\textendash reducing dissolution is inactive at lower temperatures. An increase in temperature thus shifts the balance towards the non\textendash reducing pathway \citep{PANIAS1996A}. However, this conclusion is based on the goethite ($\alpha$\textendash\,FeOOH) dissolution experiments of \citet{RUEDA1985} using aqueous ethylenediaminetetraacetic acid (EDTA); a aminopolycarboxylic acid that shares many of the dissolution mechanisms of OxA. This non\textendash reducing mechanism is illustrated on the left side of Fig.~\ref{fig:schematic_mechanisms}, where roman numerals indicate the formal oxidation states of solid iron, and arabic numerals refer to iron ions in solution. Even at elevated temperatures, the reducing mechanisms on the right side of Fig.~\ref{fig:schematic_mechanisms} are not fully suppressed. \citet{SANTAWAJA2021} for example analyzed the ferric (\ce{Fe^{3+}}) to ferrous (\ce{Fe^{2+}}) ion ratio for the dissolution of various iron oxides and found that, under dark conditions at 75$^\circ$C with 0.5~M OxA, reducing dissolution still participates to the dissolution. However, for hematite (\aFetOt) and goethite ($\mathrm{\alpha-FeOOH}$), the contribution was small and decreased with increasing molar OxA/Fe ratio.\\
While increasing temperature enhances the contribution of the non\textendash reducing pathway, the overall dissolution rate is also contingent on the concentration of \Fetwp in solution. Complexes formed with ferrous ions initiate an autocatalytic mechanism that further accelerates dissolution. Thus, a complete suppression of the reducing pathway can lead to slower overall dissolution rates. This has been shown for iron oxides that exclusively contain \Fethpc in their crystal~\citep{LAUSCH2024}. In that autocatalytic mechanism, a ferrous aqueous oxalate complex transfers an electron to the surface \ce{Fe^{III}}, and the subsequently reduced surface complex detaches from the solid~\citep{PANIAS1996A,LITTER1992}:
\begin{subequations}
\begin{align}
\left[\rangle \mathrm{Fe}^{\mathrm{III}}\mathrm{Ox}^{\mathrm{n-}}\right]+\left[\mathrm{Fe}^{2+}\,\mathrm{Ox}\right]^\mathrm{2-m}_{(\mathrm{aq})} \quad & 
\rightarrow\quad 
\rangle \mathrm{Fe}^{\mathrm{III}}\,\mathrm{Ox}^{\mathrm{n-}} \ldots \mathrm{Fe}^{2+}\,\mathrm{Ox}^\mathrm{m-} \\
\rangle \mathrm{Fe}^{\mathrm{III}}\,\mathrm{Ox}^{\mathrm{n-}} \ldots \mathrm{Fe}^{2+}\,\mathrm{Ox}^\mathrm{m-} \quad & \rightarrow\quad \rangle \mathrm{Fe}^{\mathrm{II}}\,\mathrm{Ox}^{\mathrm{n-}} \ldots \mathrm{Fe}^{3+}\,\mathrm{Ox}^{\mathrm{m-}} \\
\rangle \mathrm{Fe}^{\mathrm{II}}\,\mathrm{Ox}^{\mathrm{n-}} \ldots \mathrm{Fe}^{3+}\,\mathrm{Ox}^{\mathrm{m-}} \quad & \rightarrow\quad \rangle \mathrm{Fe}^{\mathrm{II}}\,\mathrm{Ox}^{\mathrm{n-}}+\left[\mathrm{Fe}^{3+}\,\mathrm{Ox}\right]^\mathrm{3-m}_{(\mathrm{aq})} \\
\rangle \mathrm{Fe}^{\mathrm{II}}\,\mathrm{Ox}^\mathrm{n-} \quad & \rightarrow\quad \left[\mathrm{Fe}^{2+}\,\mathrm{Ox}\right]^\mathrm{2-n}_{(\mathrm{aq})}\quad\quad\text{.}
\end{align}
\label{equ:chem_reductive_dissolution_autocatalytic}
\end{subequations}
Temperature indirectly influences the activation of the autocatalytic mechanism by altering the OxA equilibrium constants. For example, $\mathrm{pKa_1}$ and $\mathrm{pKa_2}$ shift from 1.05 and 4.26 at 25$^\circ$C to 1.73 and 4.71 at 90$^\circ$C, respectively~\citep{KETTLER1998}. As shown by \citet{LAUSCH2024}, these changes modify the distribution of iron oxalate species available for reaction (see Fig. \ref{fig:speciation_all_fe2_fe3}). Since the activation of Eq.~\ref{equ:chem_reductive_dissolution_autocatalytic} depends on the \Fetwp concentration in solution, the oxidation state of the iron oxide must be considered. Some iron oxides, such as magnetite or wüstite, naturally contain \Fetwpc. In these cases, \Fetwp is released during dissolution, leading to accelerated dissolution \citep{VEHMAANPERAE2022,SANTAWAJA2021,LITTER1992}, and the reducing pathway described in Eq. \ref{equ:chem_electron_transfer} becomes less significant. Consequently, in these cases, the effect of temperature on the balance between reducing and non\textendash reducing pathways is diminished, as the autocatalytic cycle can be activated even for primarily non\textendash reducing mechanisms at elevated temperatures. Ferrous salts may also serve as an external source of \Fetwp and have been shown to increase reaction rates \citep{TAXIARCHOU1997a,AMBIKADEVI2000,SUTER1988}. Finally, ferrous ions required for the autocatalytic cycle can be generated via an alternative light\textendash induced pathway.

Light irradiation influences dissolution based on both the light’s intensity and wavelength $\lambda$. Generally, the interaction can be classified into two different mechanisms. The first interaction mechanism is the reduction of stable aqueous iron–oxalate complexes. To determine the number of reactions a photon triggers at a specific wavelength $\lambda$, the molar absorption coefficient $\epsilon(\lambda)$ and the quantum yield $\varphi(\lambda)$ must be considered. The quantum yield describes the ratio of the number of events to the number of absorbed photons at a given wavelength over the same period of time~\citep{BRASLAVSKY2004}. An overview of reported molar absorption coefficients and quantum yields for selected aqueous iron oxalate complexes is presented in Fig.~\ref{fig:light_overview}\,(b). While ferrous oxalate complexes absorb light, particularly in the UV spectrum, the photo\textendash induced reduction is limited to ferric ions. A photo\textendash induced further reduction of \Fetwp is unlikely since $\mathrm{Fe^{1+}}$ is rarely stable and \Fetwp and \Fethp species are far more prominent \citep{MAYER2022}. Nonetheless, \citet{MAYER2022} observed a stable $\mathrm{Fe^{1+}}$ at millimolar OxA concentrations, yet its interaction with light is unclear. Additional information regarding the photoreduction of aqueous iron\textendash oxalate complexes is provided in~\ref{sec:aqueous_iron_oxalate_complexes}.\\
The second mechanism is the direct photoreduction of iron within the particle, which depends on particle properties. The crystal structure and oxidation state determine the semiconductor behavior of iron oxides via electronic states in the valence and conduction bands~\citep{LV2022}. The band gap is the energy difference between these two bands, representing the minimum energy required for an electron to transition from a bound state in the valence band to a mobile state in the conduction band. An absorption of a photon with an energy greater than the band gap energy leads to a charge separation and a corresponding hole (oxidation of the $\mathrm{O^{2-}}$) in the valence and an electron in the conduction band, which can lead to a subsequent reduction of \Fethpc to \Fetwpc \citep{SHERMAN2005}. At the surface of the iron oxides, the hole in the valence band can be refilled by the oxidation of adsorbed molecules, such as water or organic ligands~\citep{SHERMAN2005,LITTER1992}. Without ligands, the photoreduction of hematite and magnetite is unlikely due to the fast hole\textendash electron recombination~\citep{LV2022,BEYDOUN2002}. Supplementary information regarding the direct photoreduction of surface iron can be found in \ref{subsec:direct_interaction}. Ligand\textendash promoted reduction may occur when an oxalate ion bound to a surface iron is photoreduced, transferring an electron into the particle~\citep{STUMM1992}. \citet{MUKHERJEE2016} showed that immediate irradiation of the suspension produces significantly more \Fetwp than irradiating the pure liquid given the same UV light exposure, suggesting that ligand\textendash to\textendash metal charge transfer from adsorbed oxalate to surface iron plays a key role in building up the \Fetwp concentration. In EDTA, magnetite and maghemite are more easily photocorroded than hematite, which showed only negligible photoactivity \citep{LITTER1992}. However, even for magnetite and maghemite the quantum yield was estimated to be only in the order of $\mathcal{O}(10^{-5})$. \citet{LITTER1992} estimated that the reduction of aqueous complexes eventually outweighs the ligand\textendash promoted reduction of surface iron. In OxA, \citet{LITTER1991} note that the formation of the ferrous oxalate precipitates complicates the interpretation of their obtained results. Once reduced, \Fetwpc is more readily released into solution than \Fethpc \citep{SUTER1991}, and the reduced \ce{Fe^{III}} becomes even more reactive than native lattice \Fetwpc, as shown by \citet{BUXTON1983} for the dissolution of magnetite in EDTA.

Research on the dissolution of pure iron oxides has been extensive over several decades, with some of the earliest detailed investigations conducted more than 40 years ago~\citep{BAUMGARTNER1983,SELLERS1984,BLESA1987}. However, the understanding of the iron oxide dissolution remains incomplete, owing to the complexity of the interconnected mechanisms and the multitude of adjustable boundary conditions. Changes in either OxA concentration, $c_{\mathrm{OxA}}$, or temperature alter the dominant dissolution pathways and the equilibrium of iron oxalate complexes, thereby influencing both the reaction mechanisms and the dependence on short\textendash wavelength light exposure. Many experimental studies primarily focused on clarifying the fundamental dissolution mechanisms. These were performed either at temperatures below 40$^\circ$C with low $c_{\mathrm{OxA}}$ (in the mM range) \citep{SIFFERT1991,SUTER1988,MUKHERJEE2016} or at similar temperatures but with higher $c_{\mathrm{OxA}}$\,\textendash\,for example, \citet{CORNELL1987} used $c_{\mathrm{OxA}} = 0.025$ M and \citet{BAUMGARTNER1983} and \citet{BLESA1987} reached values up to 0.1\,M and 0.3\,M, respectively. Additionally, experiments with nickel chromium ferrites~\citep{SELLERS1984} promise a significant increase in reaction rates even at low $c_{\mathrm{OxA}}$ for temperatures above the boiling point (up to 140$^\circ$C), but require a pressure\textendash tight reactor. Notably, conditions of higher temperature and $c_{\mathrm{OxA}}$ remain less explored.\\
\citet{LEE2007} investigated the dissolution at $c_{\mathrm{OxA}}$ up to 0.381\,M at 100$^\circ$C for pure hematite and found that increasing $c_{\mathrm{OxA}}$ accelerates the dissolution rate, although a passivation mechanism appears to hinder the process if $c_{\mathrm{OxA}} > 0.29$ M. \citet{SALMIMIES2012} studied hematite dissolution at 35$^\circ$C and 50$^\circ$C in 0.6\,M OxA to derive a fitting rate law and found that the equilibrium concentration is unaffected by temperature. Similiarly, \citet{SALMIMIES2016,VEHMAANPERAE2021A} fitted existing rate laws for systems containing mixtures of OxA and sulfuric acid at 35 and 50$^\circ$C with $c_{\mathrm{OxA}} = 0.33$ M. Their results indicate that trace amounts of OxA enhance the solubility limit and that the measured Brunauer–Emmett–Teller specific surface area can increase again after an initial decrease during the reaction, either due to particle disintegration or formation of a solid product. In a related study, \citet{VEHMAANPERAE2022} examined the dissolution of hematite or magnetite in mixtures of nitric acid and OxA (with $c_{\mathrm{OxA}}=0.33$ M) at 35 and 50$^\circ$C. Kinetic experiments were conducted at 35$^\circ$C, while a higher temperature was used for solubility measurements; they found that nitric acid effectively prevents the formation of a solid ferrous oxalate precipitate.\\
In all of these studies at elevated temperatures and OxA concentrations, however, the influence of light remains ambiguous because the experimental setups were not explicitly characterized regarding light exposure. Given that some experiments lasted for hours, any light\textendash induced effects could have compounded over time. There are few studies at higher $c_{\mathrm{OxA}}$ and temperatures that consider light effects. \citet{SANTAWAJA2020} dissolved natural iron ores at 98$^\circ$C in 1\,M OxA, allowed the slurry to cool, and then subjected the decant to high\textendash intensity light, which produced solid ferrous oxalate products. The continuous irradiation caused the \Fetwp fraction in the decant first to increase due to the reduction of aqueous, ferric oxalate complexes until $\mathrm{Fe^{3+}/Fe^{2+}}\approx2$. It followed the precipitation of the ferrous oxalate and a corresponding decrease in both \Fetwp and \Fetwp concentration. In a subsequent study, \citet{SANTAWAJA2021} examined the dissolution of \aFetOt, FeOOH, and \FetOf at 47$^\circ$C, 75$^\circ$C, and 92$^\circ$C in the dark with $c_{\mathrm{OxA}}$ up to 1\,M. They determined the Fe solubility limit in OxA (OxA/iron molar ratio approx. 1.82) and confirmed that magnetite dissolves faster than the other oxides due to the natural presence of \Fetwpc but also reported ferrous oxalate precipitates. Other investigations have addressed additional influences. \citet{LITTER1991} studied the acceleratory effect of \Fetwp by adding ferrous salts during the dissolution of hematite and maghemite (\gFetOt) at $c_{\mathrm{OxA}} = 0.016$ M. For maghemite, a faster dissolution was observed at 70$^\circ$C without added \Fetwp. They also found that exposure to short\textendash wavelength light at 30$^\circ$C increased the reaction rate but promoted the formation of a ferrous oxalate precipitate that impeded further dissolution. \citet{TAXIARCHOU1997b} investigated hematite dissolution in 0.5\,M OxA over the temperature range 50–90$^\circ$C and demonstrated that the influence of dissolved oxygen decreases with lower pH. Furthermore, they observed that light exposure at 80$^\circ$C shortened the induction period of slow dissolution rates by enabling an additional pathway for \Fetwp generation. Lastly, \citet{LAUSCH2024} studied the combined effect of light and temperature on \FetOt particles, finding that short\textendash wavelength light’s influence diminishes with increasing temperature as the process shifts toward primarily non\textendash reducing mechanisms. At 40$^\circ$C, short\textendash wavelength light exposure led to an increase in~mean particle size, likely due to ferrous oxalate precipitation.\\
Particle properties further influence dissolution mechanisms beyond the discussed effects of oxidation state on autocatalysis and crystal structure on the photo\textendash induced reducibility of iron oxalate complexes. For example, the crystal structure of iron oxide polymorphs affects dissolution rates: spinel\textendash phase maghemite (\ce{\gamma-Fe2O3}) dissolves faster than rhombohedral hematite (\ce{\alpha-Fe2O3}) \citep{LITTER1992}. Moreover, under suppressed autocatalytic conditions at approximately 60$^\circ$C, hematite dissolves exclusively via non\textendash reducing pathways, whereas magnetite follows a reducing pathway \citep{LAUSCH2024b}. \citet{VEHMAANPERAE2021B} further suggested that iron oxalate dihydrate forms inherently during hematite dissolution. The influence of particle properties further complicates the dissolution of the CIPs. These exhibit anisotropic features—such as hollow structures and layered compositions of different iron oxides (primarily hematite and magnetite, see Sec.~\ref{subsec:Chemicals})\,\textendash\,which expose diverse surfaces and alter the dominant dissolution mechanisms. The behavior of mixed\textendash phase particles remains underexplored, as most studies in that field focus on applications such as clay purification~\citep{AMBIKADEVI2000,MARTINEZLUEVANOS2011,TAXIARCHOU1997a} or rust removal~\citep{LEE2007}, where different iron oxides coexist but lack $\mathrm{Fe}^{2+}$\textendash bearing oxides, which are present in CIPs. The effect of such mixed iron oxide dissolution was demonstrated by \citet{LEE2006} where the addition of approximately 10 wt\% magnetite can enhance the reaction rate and reduce the initial incubation period at 100$^\circ$C under controlled pH 3.0 (adjusted via ammonium hydroxide). For the dissolution of CIPs specifically, a recent systematic study in 0.5\,M OxA at around 60$^\circ$C under suppressed autocatalytic conditions concluded that the rhombohedral phase (hematite) predominantly drives the dissolution, despite the presence of magnetite~\citep{LAUSCH2024b}. However, this study was limited to a single set of temperature and light conditions, leaving their broader influence unexplored.

In summary, despite decades of research on iron oxide dissolution, significant gaps remain. The lack of systematic studies investigating the combined effects of high $c_{\mathrm{OxA}}$, elevated temperature, and controlled light exposure on reaction mechanisms for mixed\textendash phase particles motivates the present study.

\section{Materials and Methods}
\label{sec:Material_and_methods}

\subsection{Chemicals}
\label{subsec:Chemicals}
Combusted iron particles were obtained from iron powder\textendash air flames in a tube burner at a constant mean outlet velocity of 35 cm/s~\citep{FEDORYK2023} with three different fuel\textendash to\textendash air equivalence ratios $\Phi$ of 0.67, 1.0 and 1.5 (assuming full conversion to \FetOt for calculation). The analyzed particles mostly contained hematite and magnetite with traces of \aFe and \FeO of less than 3wt\% and are covered in oxide nanoparticles~\citep{BUCHHEISER2023}. While no analysis of the iron oxide phases in these nanoparticles was performed, nanoparticles recovered from the fumes during the combustion of sponge iron were shown to contain mainly maghemite (\gFetOt)~\citep{WIINIKKA2018}. A decrease in $\Phi$ also results in a higher degree of oxidation with an increase in the weight fraction, $w$, of hematite ($w(\Phi=1.5)= 0.26\pm0.03$, $w(\Phi=0.67)=0.49\pm0.03$) and a corresponding decrease in magnetite ($w(\Phi=1.5)= 0.64\pm0.03$, $w(\Phi=0.67)=0.49\pm0.03$), wüstite, and ferric iron fractions~\citep{BUCHHEISER2023}. Non\textendash combusted iron can be embedded inside oxide particles or exist as separate particles~\citep{LAUSCH2024b,DEUTSCHMANN2024}.\\
Similar to a previous study~\citep{LAUSCH2024}, the particles are used to prepare a 2.5\,g/L stock solution of the respective iron oxide. A stock solution of 5\,wt\% Pluronic F127, a nonionic surfactant (Merck, Germany), was prepared to inhibit the agglomeration of hydrophobic particles and their adhesion to the walls of the reaction vessel. Bidistilled water (CarlRoth, Germany) was employed in the preparation of all stock solutions. OxA was sourced from a 0.5\,M stock solution (Chem\textendash Lab, Belgium). The reaction was initiated by the addition of 15\,$\upmu$L of Pluronic F127 solution into the cuvette, corresponding to approximately 0.25\,wt\%, followed by 35\,$\upmu$L of bidistilled water, 2.25\,mL of OxA stock solution, and finally 200\,$\upmu$L of iron oxide suspension. This preparation yielded a total cuvette volume of 2.5\,mL. The resulting OxA concentration is $c_{\mathrm{OxA}}=\text{0.45}$\,M.

\subsection{Experimental setup and methodology}
\label{subsec:Experimental_setup}
To assess the influence of reaction temperature $\overline{T}$, light irradiation, and $\Phi$ on the dissolution rate of the CIPs in 0.45\,M~OxA, a previously developed and validated experimental setup was used~\citep{LAUSCH2024}. The setup consisted of a fused\textendash silica macro\textendash cuvette (CV10Q35FE, Thorlabs) in which a stirring motion induced by a rotating cylinder kept the particles suspended. The upper edges of the cuvette are coated with a hydrophobic material (Degussa, Tegotop 210) to prevent capillary forces from drawing particles out of the solution. The cuvette is preheated to the desired temperature using a water jacket. The fused\textendash silica cuvette transmits light from the UV range to the infrared, which allows the use of two different light sources, shown in Fig.~\ref{fig:light_overview}\,(b).

\begin{figure}[ht!]
\centering
        \begin{tikzpicture}
        \node[inner sep=0pt] at (0.1,8.7)
            {\setlength\fwidth{0.8\textwidth}
            \setlength\fheight{0.4\textwidth}
            \input{Fig_2a_complexes_absorption_and_quantum_yield}};
            \node[align=center,fill=none,draw=none] at (5.9,10.3) {(a)};
        
        \node[inner sep=0pt] at (0.0,0)
            {\setlength\fwidth{0.8\textwidth}
            \setlength\fheight{0.3\textwidth}
%
%
\definecolor{mycolor1}{rgb}{0.00000,0.51373,0.80000}%
\definecolor{mycolor2}{rgb}{0.90196,0.00000,0.10196}%
\begin{tikzpicture}[%
inner sep=0.4em, outer sep=0.25\pgflinewidth
]

\begin{axis}[%
width=0.954\fwidth,
height=\fheight,
at={(0\fwidth,0\fheight)},
scale only axis,
xmin=200,
xmax=700,
xlabel style={font=\color{white!15!black}},
separate axis lines,
every outer y axis line/.append style={mycolor1},
every y tick label/.append style={font=\color{mycolor1}},
every y tick/.append style={mycolor1},
ymode=log,
ymin=0.001,
ymax=1,
yminorticks=true,
xmin=200,
xmax=700,
xtick={\empty},
ylabel={Normalized intensity},
ylabel style={font=\color{mycolor1}},
axis background/.style={fill=white},
axis x line*=top,
axis y line*=left,
legend style={at={(0.02,0.96)}, anchor=north west, legend cell align=left, align=left, fill=none, draw=none,legend columns=3},
xmajorgrids,
ymajorgrids
]

\addplot [color=mycolor1, dashdotted, line width=2.0pt]
  table[row sep=crcr]{%
203.885096359806	0.00056521164079038\\
205.690694356921	0.000727853135303224\\
210.532979894641	0.000746704621031519\\
214.275492106845	0.000985639953261912\\
220.272266060902	0.0013543739840146\\
224.43061296335	0.00177690412772302\\
228.753105138264	0.00246218052533007\\
232.340851807857	0.00327733750096624\\
235.998946451365	0.00450482009310769\\
238.700308649647	0.00577082761968837\\
243.52617966012	0.00777768359703947\\
249.435409468862	0.00995256424901109\\
257.806818364581	0.0118639236907143\\
268.640406347275	0.0103709489943719\\
279.47399432997	0.00911112447004541\\
287.352967408293	0.0105670219677625\\
293.262197217036	0.0138945657025545\\
302.12604193015	0.0169481618853171\\
312.959629912844	0.018742254116811\\
323.793217895539	0.0196994952391708\\
334.626805878233	0.0196211490224695\\
345.460393860928	0.0216981973343984\\
356.293981843622	0.0237572527233632\\
367.127569826317	0.0272043735376205\\
377.961157809012	0.0320967609018908\\
388.794745791706	0.0376433002330032\\
399.628333774401	0.0435368354714129\\
410.461921757095	0.0505709264220223\\
421.29550973979	0.0597051551153765\\
432.129097722484	0.0738781005371116\\
442.962685705179	0.0904551620801789\\
453.796273687874	0.103993580221435\\
463.644990035778	0.128207775489228\\
471.523963114101	0.162425891264791\\
477.433192922843	0.210959484195342\\
482.357551096796	0.268475161739859\\
485.660731021556	0.326958591354165\\
490.236524175119	0.353032771332617\\
501.070112157813	0.427096732347295\\
511.903700140508	0.491999604539062\\
522.737288123203	0.556138054943781\\
533.570876105897	0.615009510784496\\
544.404464088592	0.651376686665837\\
555.238052071286	0.686238437469336\\
566.071640053981	0.734338496407353\\
576.905228036675	0.817756816557582\\
587.73881601937	0.827591762606557\\
598.572404002065	0.809650368117499\\
609.405991984759	0.806430334181468\\
620.239579967454	0.777245526356416\\
631.073167950148	0.742430178245138\\
641.906755932843	0.722006291441863\\
652.82241655177	0.725129988022806\\
657.357075913777	1\\
657.383310193835	0.879570319274192\\
660.356138706654	0.681292069057961\\
664.558803533022	0.679435734760765\\
675.392391515717	0.653543393974324\\
686.225979498411	0.618696736372235\\
697.059567481106	0.580479960285694\\
707.893155463801	0.542998513851851\\
718.726743446495	0.501901003774701\\
729.56033142919	0.468092278593315\\
740.393919411884	0.443133796098174\\
751.227507394579	0.417005973380127\\
762.061095377273	0.391637570500679\\
772.894683359968	0.349939641949298\\
783.728271342663	0.312992985017629\\
794.561859325357	0.277448304465099\\
805.395447308052	0.246676366551158\\
816.229035290746	0.221292655464123\\
827.062623273441	0.199910274918826\\
837.896211256135	0.180414137739195\\
848.72979923883	0.163469479485827\\
859.563387221525	0.148263913043381\\
870.396975204219	0.131295627638939\\
881.230563186914	0.115116721878188\\
892.064151169608	0.102860163249737\\
902.897739152303	0.090275109751932\\
913.731327134997	0.0781065223827433\\
924.564915117692	0.0655665587840035\\
932.93632401341	0.0528370050843285\\
937.532391642432	0.040248378902787\\
942.996084283055	0.0308999735476417\\
946.01686972821	0.027213387683753\\
954.111064161404	0.0310586979115769\\
964.944652144099	0.0320967609018908\\
975.778240126793	0.0284800891033591\\
986.611828109488	0.0237809326608637\\
996.952980274787	0.0207779503472833\\
};
\addlegendentry{Broadband light}
\end{axis}

\begin{axis}[%
width=0.954\fwidth,
height=\fheight,
at={(0\fwidth,0\fheight)},
scale only axis,
every outer x axis line/.append style={black},
every x tick/.append style={black},
xmin=200,
xmax=700,
xlabel={Wavelength $\lambda$ in nm},
every outer y axis line/.append style={mycolor2},
every y tick label/.append style={font=\color{mycolor2}},
every y tick/.append style={mycolor2},
ymin=0,
ymax=1.0,
ylabel style={font=\color{mycolor2}},
ylabel={Normalized intensity},
axis x line*=bottom,
axis y line*=right,
axis background/.style={fill=none},
xmajorgrids,
legend style={at={(0.02,0.9)}, anchor=north west, legend cell align=left, align=left, fill=none, draw=none,legend columns=3}
]
\addplot [color=mycolor2, dotted, line width=2.0pt]
  table[row sep=crcr]{%
555.926286140649	0.0030731681976305\\
559.350609091122	0.00353687756268205\\
562.77516025068	0.00438701139861041\\
566.199209350252	0.00438701139861041\\
569.623258449823	0.00438701139861041\\
573.047307549395	0.00438701139861041\\
576.471995634403	0.00546899991706432\\
579.897824764832	0.00848311078990051\\
589.866763331633	0.0221453626379991\\
593.608170126342	0.0324414279842403\\
597.039248065707	0.0443433016872347\\
600.005418458709	0.0596457107339421\\
602.506869577841	0.0786674553128351\\
605.195463556484	0.0983268002686741\\
606.766381358186	0.117076974316646\\
608.266257671973	0.138519238845057\\
609.558427586017	0.159642703184994\\
610.818660906555	0.185252984992331\\
611.971014087756	0.203885084896423\\
612.903088450418	0.222871407232153\\
613.863668663355	0.246202858062626\\
614.812394682195	0.271423495195162\\
615.748074114474	0.296514250769431\\
616.551223656751	0.316811196102216\\
617.368221686955	0.338595875647876\\
617.926449633082	0.361443222488446\\
618.485367911689	0.385459503353417\\
619.18461195618	0.412829089904549\\
619.82563098849	0.444095123203685\\
620.499004997628	0.471582783898697\\
621.197684224636	0.49799596988441\\
621.599319661356	0.519228062436716\\
622.235191184206	0.541777862434711\\
623.031295151356	0.572106387031449\\
623.67280229754	0.604198939337738\\
624.263618120545	0.638303475055723\\
624.915438415593	0.665897402480225\\
625.348689286815	0.69360265695467\\
625.605708137848	0.714673831316605\\
625.77234184696	0.733291762323432\\
626.046433128254	0.758172345921597\\
626.558438682761	0.784324082017292\\
626.750455110272	0.808273566652297\\
627.386622489899	0.831324338375036\\
627.649995985888	0.855626021313497\\
628.406814900236	0.881725130076492\\
628.849051504148	0.935769352503355\\
628.521588855658	0.908362656935575\\
629.86687054001	0.965038245997454\\
630.786402368397	0.987885592838023\\
632.622310219549	1\\
633.652115937559	0.974268270501817\\
633.891204074832	0.952427927431186\\
634.294709684998	0.932903186999368\\
634.745790499367	0.906089965814194\\
635.123064033146	0.871429064730251\\
635.720208601187	0.84304404054176\\
635.679458065716	0.817965092381879\\
635.979940966903	0.799687214909423\\
636.434639960233	0.779000624901837\\
636.63463914151	0.753719263923877\\
636.97751742231	0.731931548185947\\
637.121504507124	0.712202145904739\\
637.264294085899	0.690445016992651\\
637.649947877445	0.665791135750734\\
637.855846263549	0.637949252624086\\
638.253865473764	0.609134438551876\\
638.342886287597	0.584178287501629\\
638.692487390677	0.561224673931568\\
638.732217070845	0.540651435102106\\
639.027076900196	0.512852058667255\\
639.321915212692	0.485016247925149\\
639.620160360801	0.462949202498271\\
639.609215453122	0.444416284875036\\
639.950745344248	0.420345352549184\\
640.049973914195	0.400124312023178\\
640.400913843901	0.379437722015592\\
640.7136613533	0.356828210236264\\
641.305932751801	0.330649485764825\\
641.551049593501	0.306467901098424\\
642.159316184827	0.282274509017634\\
642.560608547224	0.2590020952591\\
643.219666873727	0.2329667465338\\
643.929218488638	0.204584083828186\\
644.85952030063	0.176645379903558\\
645.507616984121	0.15401056652197\\
646.214316619393	0.138368103940891\\
647.332246416318	0.118790736176374\\
648.27647780291	0.0987518671866383\\
649.221713309471	0.0804132658687588\\
651.017034990029	0.0614522451352895\\
654.518673276492	0.0391008096990109\\
657.146654668801	0.0213832783779348\\
661.76632754725	0.012888349757892\\
665.188048914163	0.00894682015495228\\
668.610044131977	0.00546899991706432\\
672.03292175825	0.00348535429989844\\
675.456499225715	0.00268674372675391\\
678.880228832569	0.00214574946752695\\
682.304095364873	0.00183660989082577\\
685.728144464445	0.00183660989082577\\
689.152193564017	0.00183660989082577\\
692.576242663589	0.00183660989082577\\
696.000109195894	0.00152747031412481\\
699.42438650455	0.00191389478500095\\
702.848207395038	0.00152747031412481\\
706.272165210976	0.00137290052577399\\
709.696442519632	0.00175932499665058\\
713.120035201036	0.000986476054897629\\
716.544084300608	0.000986476054897629\\
718.723024636699	0.000986476054897629\\
};
\addlegendentry{LED}

\end{axis}

\end{tikzpicture}%
};
            \node[align=center,fill=none,draw=none] at (5.9,2.4) {(b)};
        \end{tikzpicture}
    \caption{(a) Molar absorption coefficients for ferric and ferrous oxalate complexes along with selected experimentally determined quantum yields. Unless otherwise specified, the quantum yields refer to the $\mathrm{\left[Fe^{3+}(C_2O_4)_3\right]^{3-}}$ complex or the complex distribution is not considered. The standard deviation reported by \citet{DEMAS1981} is smaller than the used marker size. The continuous fit provided by \citet{STRAUB2018} is based on experimental data from refs. \citep{HATCHARD1956,PARKER1959,GOLDSTEIN2008}. (b) Relative intensities, as specified by the manufacturer, for the combined Deuterium/Halogen lamp (Ocean Optics, DH-2000-S-DUV) used for continuous broadband light irradiation and the pulsed LED (ILA5150 LPSv3) used for particle illumination.}
    \label{fig:light_overview}
\end{figure}

One light source\,\textendash\,a combined Deuterium/Halogen lamp (Ocean Optics, DH-2000-S-DUV)\,\textendash\,provides a broadband spectrum between 190 and 2500 nm with a nominal output power of 585 $\upmu$W/990 $\upmu$W, respectively. A fused\textendash silica window is installed in the water jacket to prevent UV absorption by the polymethyl methacrylate (PMMA) of the jacket. This light source is used to examine the influence of short\textendash wavelength light. The second light source is a pulsed LED (ILA5150 LPSv3, peak 630-640 nm) employed for particle illumination as part of a video system. Because both the quantum yield and the molar absorption coefficients tend to zero at $\lambda > 550$ nm (shown in Fig.~\ref{fig:light_overview}\,(a)), the LED used for particle illumination (shown in Fig.~\ref{fig:light_overview}\,(b)) is assumed to have a negligible effect on photo\textendash induced reduction.
 
\begin{figure}[!htbp]
    \centering
    \definecolor{Emerald}{rgb}{0.31, 0.78, 0.47}
    \begin{tikzpicture}
    \node[inner sep=0pt] (main_picture) at (0,0) {\includegraphics[width=0.6\textwidth]{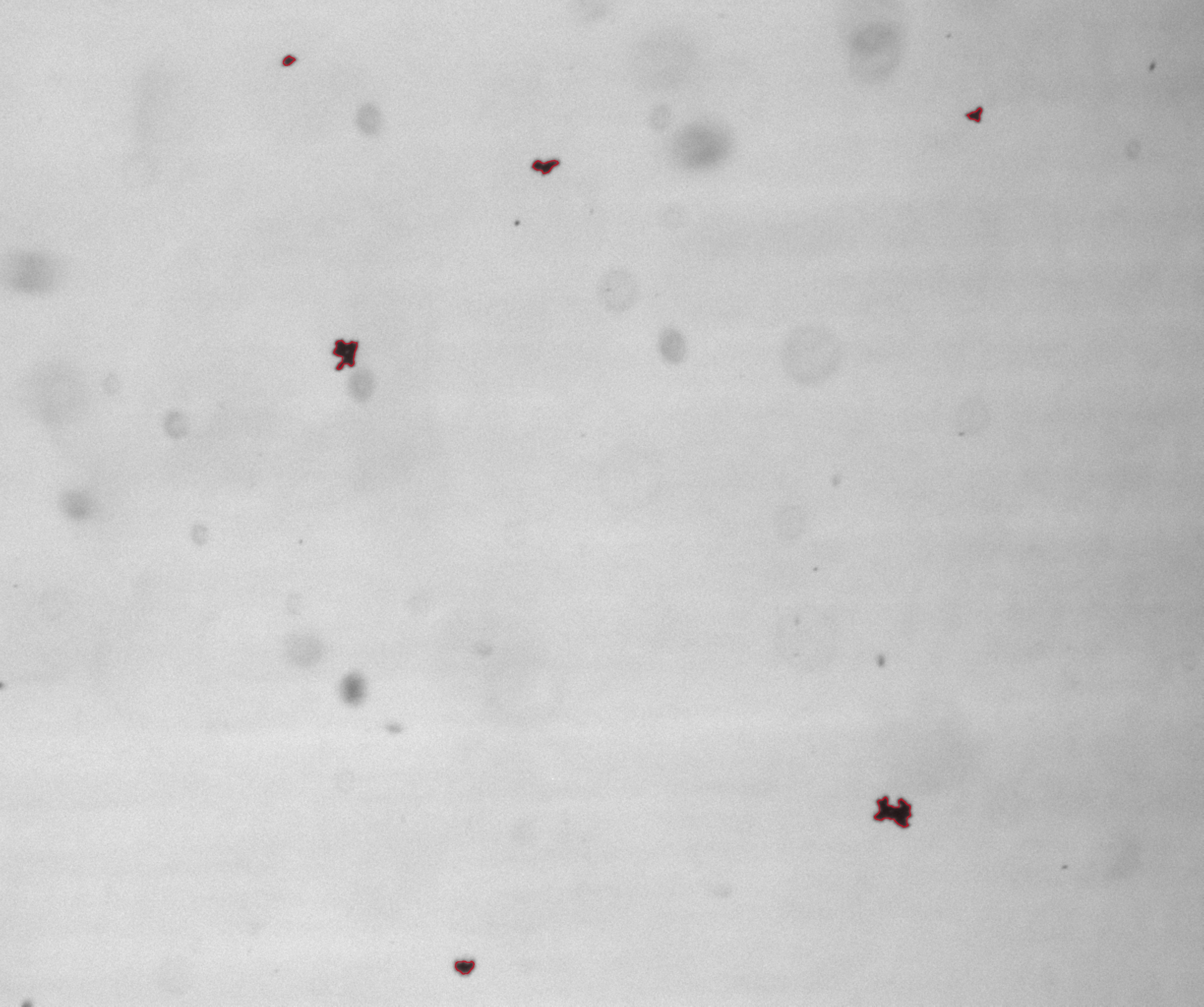}};
        \node[align=center,fill=white,draw=none,fill opacity=0.4,text opacity=1] at (-5.0,4.16) {(a)};
        \draw [draw=none,fill=white,fill opacity=0.4] (-0.5,-2.0) rectangle (5.0,1.8);
        \node[anchor=south west, align=left] at (-0.25,-0.85) {\includegraphics[width=0.12\textwidth]{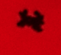}};
        \node[anchor=south west, align=left] at (2.4,-0.85) {\includegraphics[trim=1740 317 560 1596,clip,width=0.12\textwidth]{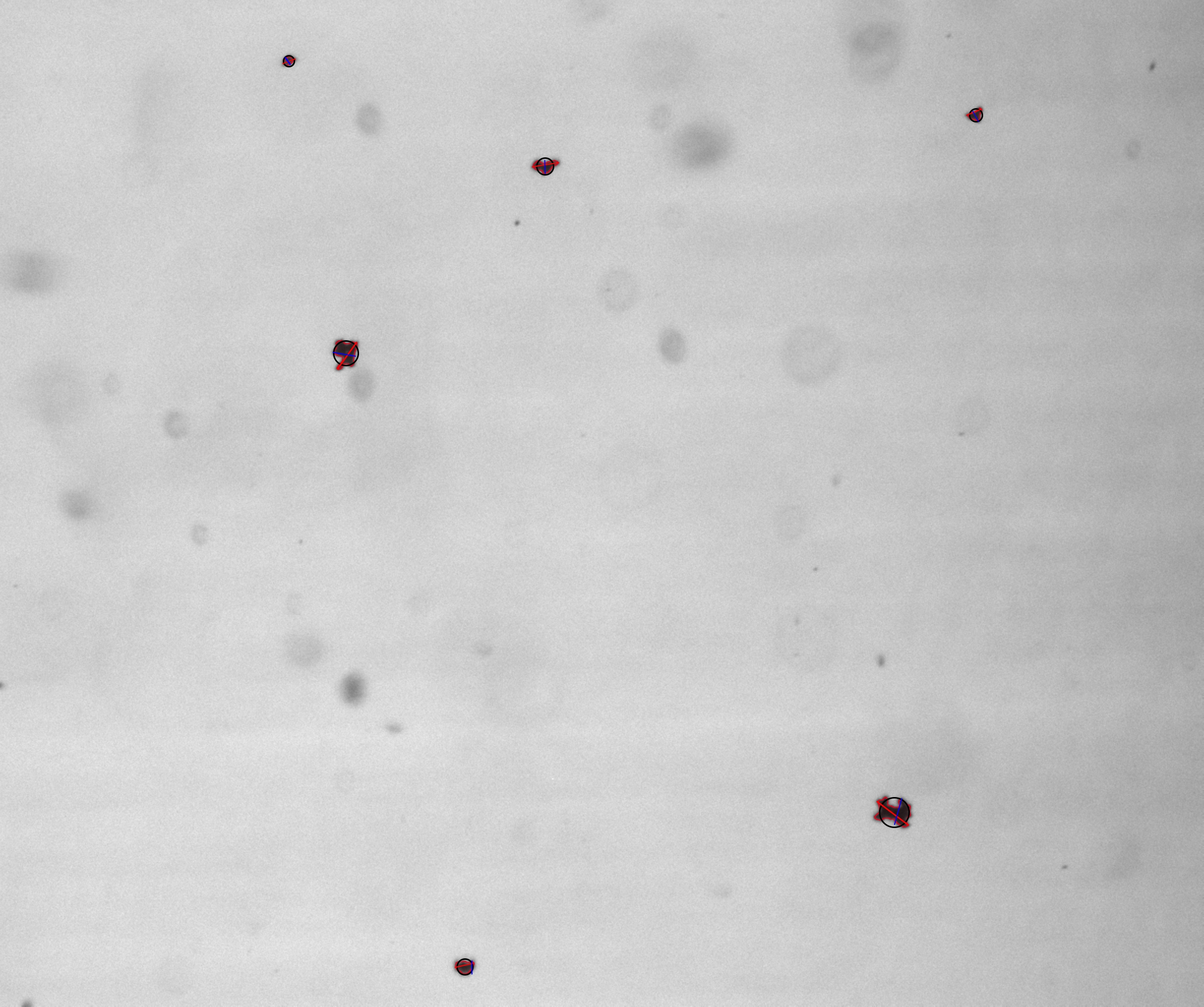}};
        \draw[line width=0.8mm,color=black] (0.18,-0.5) -- (1.7,-0.5);
        \node[anchor=south west, align=left,fill=none,draw=none] at (0.38,-0.55) {50\,$\upmu$m};
        \node[align=center,fill=none,draw=none] at (-0.2,1.5) {(b)};
        \draw[line width=0.4mm,color=Emerald] (3.65,0.04) -- (3.87,0.82);
        \node[align=center,anchor=south west] at (-0.3, -1.95) {$d=\sqrt{4 A/\pi}\approx24.1\,\upmu$m\\
        \textcolor{red}{$\mathrm{F_{max}}\approx33.1\,\upmu$m},  \textcolor{Emerald}{$\mathrm{F_{min}}\approx22.9\,\upmu$m}};
        \draw[white,line width=2pt,draw opacity=0.4] (2.5,-2.4) -- (2.25,-1.98);
        \node[anchor=south west, align=left,fill=none,draw=none] at (3.85,-4.24) {100\,$\upmu$m};
        \draw[line width=0.8mm,color=black] (4.03,-4.2) -- (5.08,-4.2);
        \node[anchor=south west, align=left,fill=none,draw=none] at (-5.3,-4.5) {$t=3921.0$\,s\\$\overline{T}=59.9^\circ$C\\$\Omega=1009.6$\,rpm};
        \node[anchor=north west, align=left,fill=none,draw=none] at (-1.15,-3.8) {13.3\,$\upmu$m};
        \node[anchor=north west, align=left,fill=none,draw=none] at (-3.81,1.57) {19.9\,$\upmu$m};
        \node[anchor=north west, align=left,fill=none,draw=none] at (-0.36,3.25) {14.0\,$\upmu$m};
        \node[anchor=north west, align=left,fill=none,draw=none] at (3.38,3.68) {10.3\,$\upmu$m};
        \node[anchor=north west, align=left,fill=none,draw=none] at (-2.7,4.15) {8.7\,$\upmu$m};
        
    \end{tikzpicture}
    \caption{(a) Example image taken from an experimental run ($\Phi=1.0$, $\overline{T}=60^\circ$C, with broadband light) approximately 65\,min after reaction begin. Red contour lines indicate the detected object perimeter, the number next to each object denote the area\textendash equivalent particle diameter $d$. (b) Unprocessed image and the corresponding processed image of an exemplary detected object. Indicated are the black area\textendash equivalent circle at the centroid of the object and the minimum and maximum Feret diameter in green and red ($F\approx1.45$).}
\label{fig:example_image_processing}
\end{figure}

At the start of the experiments ($t=0$), CIPs were added to the solution, and cylinder rotation and image acquisition (using a IDS Imaging
U3-3080CP-C-HQ video camera, spatial resolution 0.4 $\upmu$m/pixel) were immediately activated. An experiment was considered complete when no particles were visually detected (the optical resolution limit was intentionally set at 5\,$\upmu$m during the postprocessing, corresponding to the area\textendash equivalent particle diameter) or when the reaction time reached 10\,h. The cylinder was then cleaned in an ultrasonic water bath for 3\,min and rinsed with bi\textendash distilled water. Finally, the hydrophobic coating on the top edges was renewed.\\
Post\textendash processing of the data obtained from the video system under different boundary conditions yielded the projected area $A$ of each detected particle and its Feret diameter ratio $F = F_\mathrm{max}/F_\mathrm{min}$ (where $F=1$ corresponds to a circle and higher values indicate needle\textendash shaped objects). The geometric descriptors are presented in Fig.~\ref{fig:example_image_processing}. The values of $A$ obtained from the recorded images of all particles were then used to calculate the area\textendash equivalent particle diameter $d = \sqrt{4A/\pi}$. The total number of particles $n$ in each of the 60 discrete time intervals was then used to estimate the median particle diameter $\Tilde{d}$. A bootstrap procedure with 4000 bootstrap samples yielded an estimation of the 95\% confidence interval of the non\textendash normal distribution $\Tilde{d}_{95,\mathrm{boot}}$ in each time interval~\citep{LAUSCH2024}.

The discrete values (levels: such as 40$^\circ$C) of the controlled parameters (factors such as reaction temperature $\overline{T}$) were selected such that they were consistent with previous experiments, and a significant size effect was expected, defined as the difference between the mean level values per variance~\citep{ERCEG2008}. The equidistant\textendash level distribution also complied with a full factorial design, which allowed the determination of the significance of each factor and resolved the interactions between the factors without confounding. To determine the significance of a factor, the mean of each level was first calculated using all experimental runs with that level, irrespective of other factor\textendash level conditions. Because of the symmetric level distribution of the experimental runs, the influence of all other factors on the mean was eliminated  (orthogonality). For single\textendash factor and two\textendash level conditions, the significance can be determined using a simple two\textendash sample t\textendash test. In the case of three systematically varying parameters as in the present study, three\textendash way analysis of variance (ANOVA) was performed. The discrete values of the parameters used in this study are summarized in Tb.~\ref{tb:Factor_levels_reaction}.

\begin{table*}[!htbp]
	\begin{tabular*}{\textwidth}{@{}l @{\extracolsep{\fill}} ccc@{}}
		\toprule[1.5pt]
		\textsc{\textbf{Controlled parameters}} & \textsc{\textbf{Level 1}} & \textsc{\textbf{Level 2}} & \textsc{\textbf{Level 3}}\\
		\midrule
		Stirring speed in rpm & 1000 & & \\
        \rule{0pt}{1.3ex}Particle loading in g/L & 0.2 & &  \\
        \rule{0pt}{1.3ex}OxA concentration $c_{\mathrm{OxA}}$ in M & 0.45 & & \\
        \rule{0pt}{1.3ex}Light wavelength $\lambda$ in nm & 630-640 (peak) & \specialcellc{630-640 + \\[1pt]190-2500 (broadband)}  &  \\
		\rule{0pt}{1.3ex}Fuel\textendash to\textendash air equivalence ratio $\Phi$ of CIPs & 0.67 & 1.0 & 1.5 \\
		\rule{0pt}{1.3ex}Reaction temperature $\overline{T}$ in $^\circ$C & 40 & 60 & 80\\
		\bottomrule[1.5pt]
	\end{tabular*}
	\caption{Mean values of experimental parameters varied during experiments. For convenience, the light irradiation level values are denoted \textit{no broadband light} (\textbf{Level 1}) and \textit{broadband light} (\textbf{Level 2}) in the manuscript.}
	\label{tb:Factor_levels_reaction}
\end{table*}

There were 18 possible combinations of these sets of parameters. Each combination of parameters was repeated twice, resulting in 36 experimental runs. In addition, three experiments were conducted at approximately 25$^\circ$C without broadband light. Despite our best efforts to set $\overline{T}$ exactly at the specified level, variations of approximately~$ \pm 0.35$\,K (with a maximum deviation of 1.5\,K) from the target temperature were observed. The experimental $\overline{T}$ values are presented in Fig.~\ref{fig:fit_rates}. Thus, the $\Phi$ and light irradiation factors were treated categorically in the statistical analysis, and $\overline{T}$ was assumed to be a continuous variable. However, treating $\overline{T}$ also categorically does not change the outcome of the statistical analysis in Sec.~\ref{subsec:determination_of_significant_factors}.

\section{Results and Discussion}
\label{sec:results_and_discussion}
The evolution of the median particle diameter for all the combinations of temperature, light irradiation, and $\Phi$ is shown in Fig.~\ref{fig:raw_data_all}. As the particles dissolve, their average size decreases, as observed from the evolution of $\Tilde{d}(t)$. At the beginning of the experiment, $\Tilde{d}$ and the maximum $\Tilde{d}_{\mathrm{max}}$ varies between approximately 7.5\,$\upmu$m and 9.7\,$\upmu$m. In agreement with the literature~\citep{BUCHHEISER2023,LAUSCH2024b}, $\Tilde{d}_{\mathrm{max}}$ increases slightly with increasing $\Phi$ from approximately 8.1\,$\upmu$m ($\Phi=0.67$), to 8.3\,$\upmu$m ($\Phi=1.0$), and finally 8.5\,$\upmu$m ($\Phi=1.5$). However, as shown in Fig.~\ref{fig:oxs_diameter}, the scatter in the data is high, and subsequent one\textendash way ANOVA reveals that this trend is insignificant at the $\alpha=5\%$ significance level. \\
Changing the boundary conditions, as specified in Tb.~\ref{tb:Factor_levels_reaction}, results in different dissolution rates. In this section, the main trends and significant parameters for dissolution are identified, and the observed evolution in particle morphology is discussed. Subsequently, a detailed investigation of the influence and interaction of significant factors is presented, and recommendations for optimizing future dissolution processes are provided.  

\subsection{Variance analysis of the influencing parameters}
\label{subsec:determination_of_significant_factors}
\begin{figure}[ht!]
\centering
    \begin{minipage}[b]{0.4\textwidth}
        \begin{tikzpicture}
        	\node[inner sep=0pt] at (0,0)
        	{\includegraphics{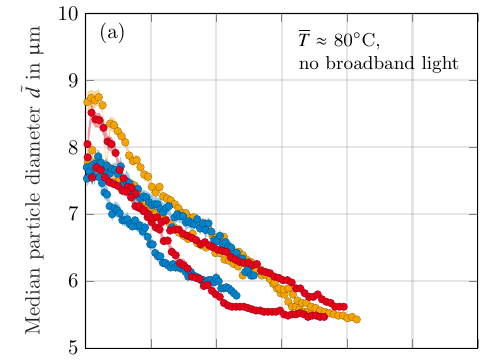}};
        \end{tikzpicture}
    \end{minipage}  
    \hspace{0.85cm}
    \begin{minipage}[b]{0.4\textwidth}
    	\begin{tikzpicture}
    		\node[inner sep=0pt] at (0,0)
    		{\includegraphics{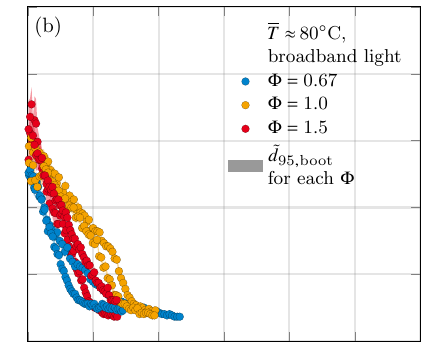}};
    	\end{tikzpicture}
    \end{minipage}
    
    \begin{minipage}[b]{0.4\textwidth}
    	\begin{tikzpicture}
    		\node[inner sep=0pt] at (0,0)
    		{\includegraphics{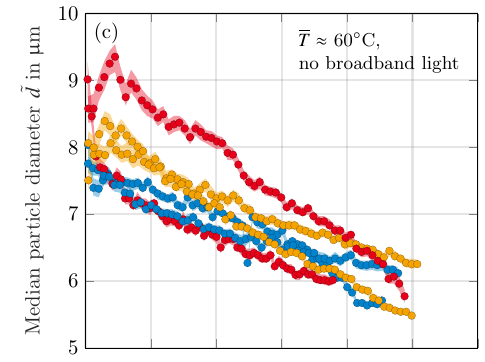}};
    	\end{tikzpicture}
    \end{minipage}
    \hspace{0.85cm}
    \begin{minipage}[b]{0.4\textwidth}
    	\begin{tikzpicture}
    		\node[inner sep=0pt] at (0,0)
    		{\includegraphics{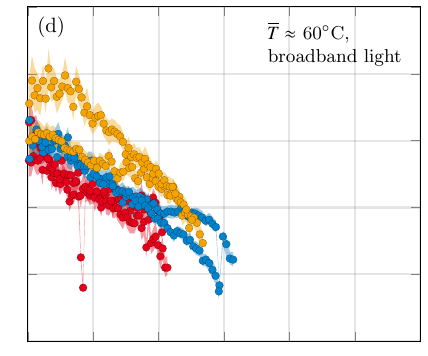}};
    	\end{tikzpicture}
    \end{minipage}
    
    \begin{minipage}[b]{0.4\textwidth}
    	\begin{tikzpicture}
    		\node[inner sep=0pt] at (0,0)
    		{\includegraphics{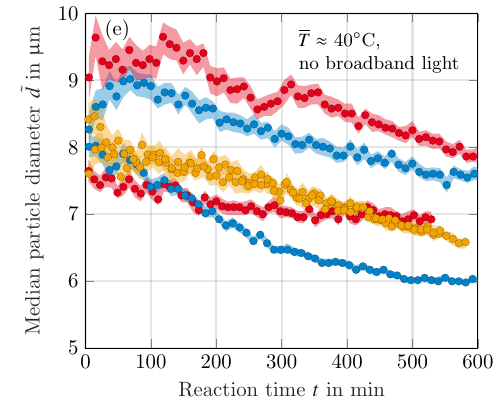}};
    	\end{tikzpicture}
    \end{minipage}
    \hspace{0.85cm}
    \begin{minipage}[b]{0.4\textwidth}
    	\begin{tikzpicture}
    		\node[inner sep=0pt] at (0,0)
    		{\includegraphics{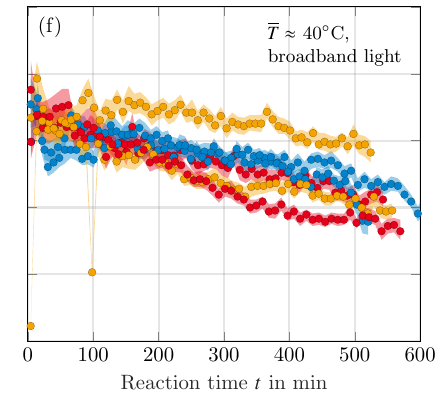}};
    	\end{tikzpicture}
    \end{minipage}
    \caption{Median particle diameter $\Tilde{d}(t)$ at $\overline{T}=80^\circ$C (a,b), 60$^\circ$C (c,d) and 40$^\circ$C (e,f), both without (a,c,e) and with (b,d,f) broadband light irradiation for varying $\Phi$ (0.67, 1.0, 1.5), each measured two times.}
    \label{fig:raw_data_all}
\end{figure}

To identify the significant parameters influencing the dissolution process, the rate of change in particle size, $C:=\mathrm{d}\Tilde{d}/\mathrm{d}t$, was calculated. Subsequently, the significance of the factors outlined in Tb.~\ref{tb:Factor_levels_reaction} was evaluated using three\textendash way ANOVA. In most CIP dissolution experiments, $C$ remains almost constant. However, closer inspection of the data presented in Fig.~\ref{fig:raw_data_all} shows that most experimental runs exhibit a deviation from a linear trend in the early stages. This is partially attributed to the gradual suspension of the particles as the spinning cylinder accelerates at the start of the experiment and clusters of particles start to break up, as discussed in detail in Sec.~\ref{subsec:particle_morphology}. The influence of gradual suspension was expected to be greater for experiments with shorter durations, owing to the shorter elapsed time per interval. In addition to these initial fluctuations, a distinct change in $C$ is observed towards the end of some experimental runs, particularly those performed at higher temperatures (Fig.~\ref{fig:raw_data_all}\,(a) and (b)). The evolution of $\Tilde{d}(t)$ contrasts with that observed for mixed\textendash phase \FetOt particles~\citep{LAUSCH2024}, which exhibited a distinct change in $C$ at the beginning of the experiment due to a shift from a breakup\textendash dominated phase to a dissolution\textendash dominated phase.  

\begin{figure}[ht!]
\centering
    \begin{minipage}[b]{0.45\textwidth}
    	\begin{tikzpicture}
    		\node[inner sep=0pt] at (0,0)
    		{\includegraphics{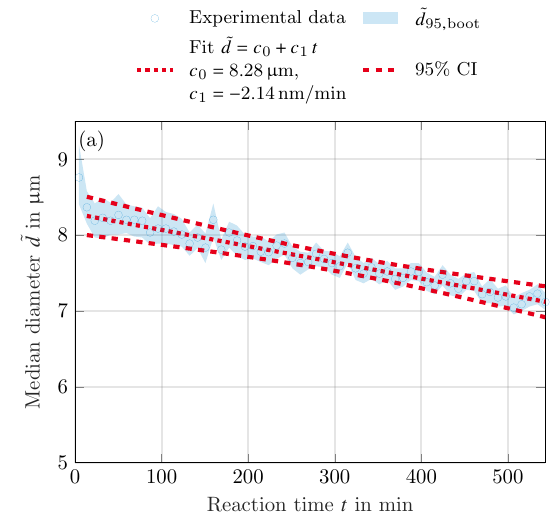}};
    	\end{tikzpicture}
    \end{minipage}    
    \hspace{1cm}
    \begin{minipage}[b]{0.45\textwidth}
    	\begin{tikzpicture}
    		\node[inner sep=0pt] at (0,0)
    		{\includegraphics{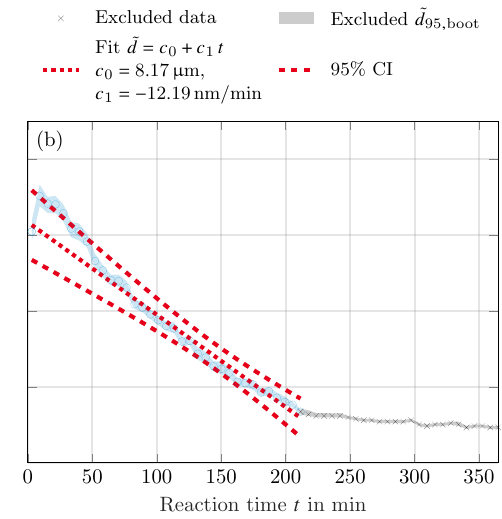}};
    	\end{tikzpicture}
    \end{minipage}
    \caption{(a) Linear fit for the factor combination $\overline{T}\approx40.0^\circ$C, with broadband light, $\Phi=0.67$, with 95\% CI$(c_0)= [8.1,\,8.5]\,\upmu$m, CI$(c_1) = [1.5,\,2.7]$\,nm/min. (b) Segmented linear fit for the factor combination $\overline{T}\approx80.2^\circ$C, without broadband light, $\Phi=0.67$, with 95\% CI$(c_0)= [7.8,\,8.5]\,\upmu$m, CI$(c_1) = [9.8,\,14.6]$\,nm/min.}
    \label{fig:fit_example}
\end{figure}

The dissolution rate was determined from the data presented in Fig.~\ref{fig:raw_data_all}; the initial $\Tilde{d}(t)$ values were outlier-filtered and the data linearly fitted using a robust iterative procedure based on the Huber loss function (MATLAB fitlm)~\citep{HUBER1964}. The weights were taken as the inverse variance derived from the $\Tilde{d}_{\mathrm{95,boot}}(t)$ values. Thus, both the outliers and $\Tilde{d}(t)$ values with higher uncertainty contributed less to the fit. To account for the change in $C$ observed in some of the experimental runs, the experimental data were split into two segments based on the inflection point of the first derivative of the data. Thus, for certain experimental runs with segmented data, two rate values were obtained, denoted as $c_1$ and $c_2$. In subsequent statistical analyses, only the initial rate $c_1$ was used. However, for a detailed discussion on the influence of temperature and light on the dissolution process in Sec.~\ref{subsec:temp_and_light_influence}, both $c_1$ and $c_2$ are considered.\\
The fit shown in Fig.~\ref{fig:fit_example}\,(a) provides an example of non\textendash segmented experimental data. Although most $\Tilde{d}$ values closely follow a linear trend, there is a deviation in the initial $\Tilde{d}(t\approx13.7\,\mathrm{min})$, where the initial particle suspension process is expected to exert the greatest influence. Fig.~\ref{fig:fit_example}\,(b) shows a representative fit of segmented data. In this case, $\Tilde{d}(t)$ deviates from the ideal linear trend, as reflected by the $>95$\% confidence interval (CI) of the fit. Although it may be tempting to use higher\textendash order polynomials to fit these experimental data, some of these models also do not fit the data well, suggesting that increased complexity does not necessarily yield a better model. Therefore, a simple first\textendash order polynomial was deemed more appropriate for the data presented in Fig.~\ref{fig:fit_example}\,(b). The inaccuracies introduced using a linear fit were mitigated by including the 95\% CI of the fit parameters in the analysis.

\begin{figure}[ht!]
    \centering
        \begin{tikzpicture}
        	\node[inner sep=0pt] at (0,0)
        	{\includegraphics{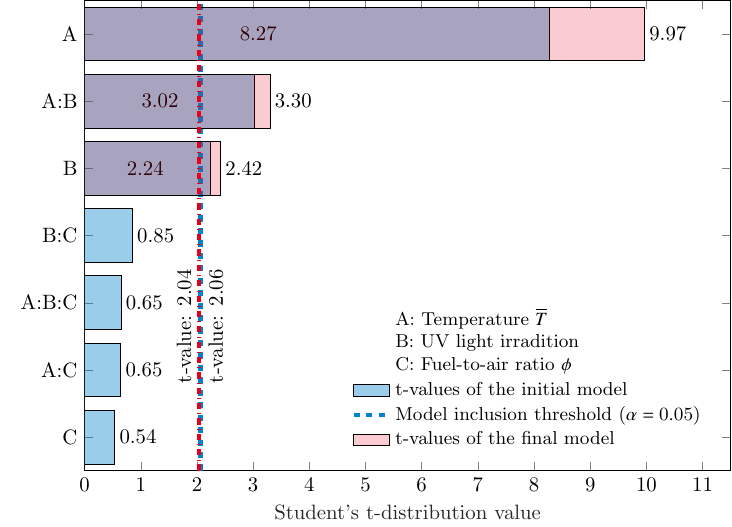}};
        \end{tikzpicture}
    \caption{Graphical representation of the ANOVA results in form of a Pareto chart.}
    \label{fig:pareto_chart_significant_values}
\end{figure}

The $c_1$ values extracted from the fitting procedure were used to determine the significant factors. The significance of an effect or interaction effect on the measured response was determined by performing an ANOVA with a null hypothesis to prove that there was no effect on $c_1$ at the $\alpha=5\%$ threshold. ANOVA requires mutual independence, normal distribution, and equal variances of the error variable for any inference to be valid. In addition to the inspection of factor\textendash level variance, this can be tested by examining the standardized residuals. Standardized residuals are obtained by calculating the difference between the measured $c_1$ and the value predicted by ANOVA for that factor\textendash level combination divided by its estimated standard deviation~\citep[pp.143-144]{DEAN2017}. As described in~\ref{sec:appendix_statistical_analysis}, the violation of some requirements necessitated $c_1$ data transformation. The subsequently obtained ANOVA model for the transformed rate $\Xi$ includes insignificant terms that must be excluded systematically. This was achieved through an adjusted backward elimination procedure in which the term with the highest p\textendash value (most insignificant) was iteratively excluded, and the ANOVA results were recalculated using the reduced model. After each step, if the reintroduction of any excluded model term results in a p\textendash value below the $\alpha=5\%$ threshold, the term is included in the next model step and ANOVA is repeated. After completing the procedure, only significant terms remain in the model, and the standardized residuals are again inspected. After this procedure, normality, homoscedasticity, and independence were all deemed suitably met, as shown in Figs.~\ref{fig:residuals_normality},~\ref{fig:residuals_fitvals}, and \ref{fig:residuals_runtime}. The significance of each term in the initial and reduced ANOVA models is graphically presented in blue and red, respectively, in Fig.~\ref{fig:pareto_chart_significant_values}. The null hypothesis was rejected and an effect was considered significant when the standardized effect strength, represented by the t\textendash value defined in Eq.~\ref{equ:pareto_values}, was above the threshold associated with a global significance level of 5\%. Note that this threshold is dependent on the degrees of freedom of the ANOVA model, and thus varies slightly depending on the number of terms included in the model.
Following a previously reported method~\citep{LAUSCH2023}, the t\textendash values in Fig.~\ref{fig:pareto_chart_significant_values} were calculated as the inverse cumulative distribution function (CDF) of the Student's t\textendash distribution for a given ANOVA p\textendash value $p_\mathrm{A}$ as: 
\begin{equation}
\text{t\textendash value} = F_{t}^{-1}\left(\left.1-\frac{p_{\mathrm{A}}}{2}\,\right\vert \,\mathrm{DF}_{\mathrm{E}}\right)\label{equ:pareto_values}\text{,}
\end{equation}
where $F_t^{-1}$ denotes the inverse Student's t CDF and $\mathrm{DF_E}$ denotes the degrees of freedom of the error. For terms with more than one degree of freedom (e.g., $\Phi$), $\mathrm{DF_E}$ equals the total degrees of freedom of the model (number of experiments minus one) minus the sum of the degrees of freedom of each included term. If a term has only one degree of freedom (e.g., light irradiation), then $\mathrm{DF_E}$ equals the total number of degrees of freedom of the model minus two, thus reverting to a simple two\textendash sample t\textendash test.

In addition to the significant effects of $\overline{T}$ and light, which were deduced from the raw data in Fig.~\ref{fig:raw_data_all}, ANOVA reveals an interaction between the factors. Therefore, the effect of temperature on $c_1$ depends on the state of light irradiation. Interestingly, the results presented in Fig.~\ref{fig:pareto_chart_significant_values} also indicate that $\Phi$ does not affect $\Xi$, nor does it participate in any significant interactions (see Fig.~\ref{fig:main_effects_oxs}). This is surprising, considering that increasing $\Phi$ generally increases the \Fetwpc content, which is expected to enhance the autocatalytic pathway (Eq.~\ref{equ:chem_reductive_dissolution_autocatalytic}) and thus affect $\Xi$. These findings indicate that $\Phi$ does not affect the dissolution mechanisms for reaction times exceeding the three\textendash hour period observed in previous X\textendash ray absorption spectroscopy measurements~\citep{LAUSCH2024b}, even with the activation of the autocatalytic mechanism in the current work compared to the previous study. The observed lack of influence of $\Phi$ may be because the effect is too small to be detected using the current method or could be due to hematite interstitials blocking access to the magnetite core regardless of the fraction of magnetite in the particle. In the latter case, ferrous ions can still influence the dissolution process once the hematite shell is dissolved (see Sec.~\ref{subsec:temp_and_light_influence}). However, if hematite interstitials block the access to the magnetite core, the influence of the increased reaction rate of spinel\textendash type magnetite is suppressed. This decreases the differences in $C$ between the various $\Phi$ conditions. Previous studies~\citep{BUCHHEISER2023,LAUSCH2024b} showed that $\Tilde{d}$ is a function of $\Phi$ and a slight dependency is observed in the experimental data in Fig.~\ref{fig:oxs_diameter}. Therefore, the potential correlation between $\Tilde{d}_{\mathrm{max}}$ and $\Xi$ was investigated, because it could obscure the potential influence of $\Phi$. This effect was analyzed by plotting the standardized residuals against $\Tilde{d}_{\mathrm{max}}$. As shown in Fig.~\ref{fig:residuals_diam}, there is no discernible trend, indicating that $\Tilde{d}_{\mathrm{max}}$ has no significant effect on the statistical analysis presented in Fig.~\ref{fig:pareto_chart_significant_values}.

\subsection{Particle clustering}
\label{subsec:particle_morphology}
\begin{figure*}[ht!]
    \centering
    \begin{tikzpicture}
    	\node[inner sep=0pt,anchor=north west,align=left] at (0,0){\includegraphics{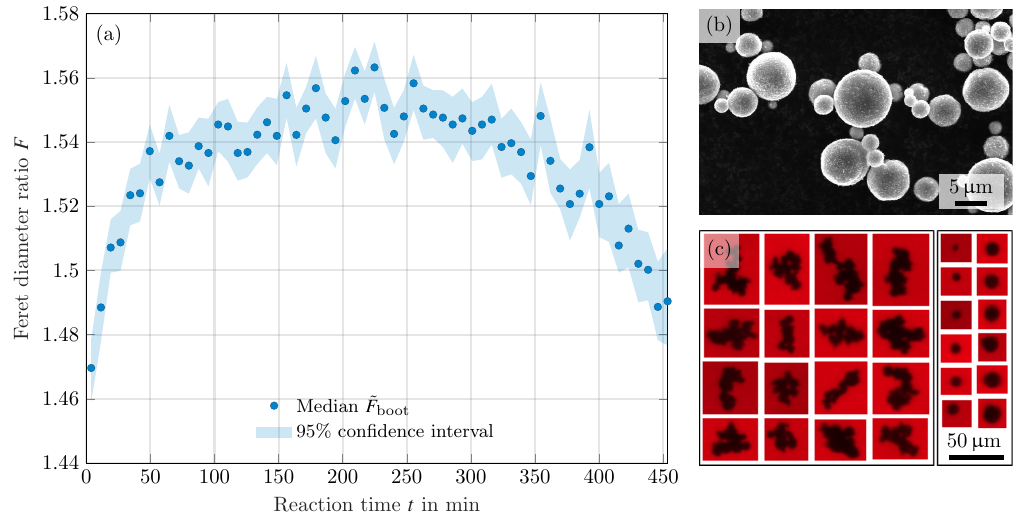}};
    \end{tikzpicture}
    \caption{(a) Evolution of the Feret diameter ratio at $\overline{T}\approx 59.9^\circ$C, without light, and $\Phi=1.5$. Median $\Tilde{F}\approx1.54$ with a 95\% CI of [1.52, 1.55]. (b) SEM images of CIPs with \POnepf. (c) Representative images of the particles and clusters detected in this study.}
    \label{fig:particle_images_clusters_SEM}
\end{figure*}

The detected objects in the recorded images exhibit various morphologies, which were generally non\textendash spherical, consistent with the $\Tilde{F}>1$ values in Fig.~\ref{fig:particle_images_clusters_SEM}\,(a). In general, the median $\Tilde{F}\approx1.54$ (95\% CI [1.50, 1.55]) for all experiments indicates elongated or irregularly shaped particles. These irregular shapes observed in the recorded images do not agree with the scanning electron microscopy images of spherical CIPs in Fig.~\ref{fig:particle_images_clusters_SEM}\,(b), which is consistent with various reports in the literature~\citep{WIINIKKA2018,TOTH2020,Li2021,CHOISEZ2022,POLETAEV2022}. This implies that most of the detected objects are clusters of spherical particles rather than single non\textendash spherical particles, as shown in Fig.~\ref{fig:particle_images_clusters_SEM}\,(c). However, as shown on the right of Fig.~\ref{fig:particle_images_clusters_SEM}\,(c), some objects with almost perfectly spherical morphologies are observed, which are probably large single particles. Although the particles clustered in solution, no residue was observed on the cuvette walls during or after the experiment, owing to the use of the F127 surfactant. The evolution of $\Tilde{F}$ over time suggests that different fragmentation mechanisms occur during dissolution. While $\Tilde{F}$ fell continuously for \FetOt particles in a previous study \citep{LAUSCH2024}, indicating their gradual breakup, the detected CIPs in the present study first became more needle\textendash shaped before transitioning toward rounder shapes.

The observed clustering of the particles is attributed to the experimental methods. As described in Sec.~\ref{subsec:Experimental_setup}, owing to the small sample weights used in the experiment, the particles were suspended in water before they were introduced into the setup. Large particle clusters were observed in the dry by SEM measurements and during the sample preparation in water.
The evolution of the Feret diameter ratio of these clusters also depends on the flow state of the system. The experimental configuration\,\textendash\,a rotating cylinder in a cuvette\,\textendash\,is a variant of a Taylor\textendash Couette system that employs a square outer container rather than a concentric cylinder. The flow in a such a system has rarely been studied~\citep{SIERRA2017}. Unlike the classical case of concentric cylinders, this configuration breaks the axis symmetry, leading to a distinct flow state characterized by additional vortices in the corners of the container. According to \citet{ESSER1996} and \citet[pp.272–324]{CHANDRASEKHAR2013}, toroidal (Taylor) vortices appear when it becomes energetically favorable for concentric fluid rings to exchange positions. The onset of the centrifugal instability occurs when the Reynolds number $\mathrm{Re}$ surpasses a critical threshold and small disturbances in the flow are amplified rather than damped. This centrifugal instability then ultimately leads to the formation of the toroidal Taylor\textendash vortices. For a fixed outer cylinder, the Reynolds number is defined as~\citep{ESSER1996}:

\begin{equation} 
\mathrm{Re} = \frac{(r_\mathrm{o}-r_\mathrm{i})\,\Omega r_\mathrm{i}}{\nu}\text{\,, }
\label{equ:TC_Reynoldsnumber} 
\end{equation}

where $r_\mathrm{i}$ and $r_\mathrm{o}$ represent the inner and outer cylinder radii, $\Omega$ is the angular velocity, and $\nu$ is the kinematic viscosity. In the square container variant, the flow remains qualitatively similar to the classical case, prompting \citet{SNYDER1968} and \citet{SIERRA2017} to use the Reynolds number for the smallest gap as an approximate indicator of the first bifurcation producing Taylor vortices. Following their approach and using the equation by \citet{ESSER1996}, the critical Reynolds number for the present setup is $\mathrm{Re_c}\approx73$, corresponding to a rotational speed of $\approx117$\,rpm (assuming 20$^\circ$C viscosity values from \citep[pp.762–763]{BAER2019}). At the operating speed of 1000\,rpm, $\mathrm{Re}\approx626$ per Eq. \ref{equ:TC_Reynoldsnumber}. Depending on the radius ratio of the inner and outer cylinder $\eta:=r_\mathrm{i}/r_\mathrm{o}$, another critical Reynolds number would indicate a transition to a flow state with azimuthal waves superimposed on the vortices that propagate with the flow~\citep{DIPRIMA1984,GROSSMANN2016}; however, the wide gap (maximum $\eta=0.6$) in the present system precludes this instability~\citep{DIPRIMA1984}. Moreover, the flow remains below the threshold for turbulence onset at $\mathrm{Re}\approx1218$ \citep{GROSSMANN2016}.\\
Taylor vortices influence the fragmentation and morphology of the clusters, contributing to the evolution seen in Fig.~\ref{fig:particle_images_clusters_SEM}. For example, \citet{RUAN2020} employed discrete\textendash element method computations to investigate fragmentation mechanisms for dense agglomerates in simple shear flows and Taylor\textendash Green vortices without wall effects. In both flow types, the agglomerates first stretched into a chain\textendash like structure, increasing the aspect ratio of the cluster. The fluid stress then surpassed the weakened connections, causing an initial breakage that triggered a cascade of further fragmentation until a steady\textendash state size distribution was reached that consisted of clusters of smaller aspect ratio compared to the chain\textendash like structure. Although the fragmentation mechanism remained consistent between vortex and shear flow conditions, the location\textendash dependent shear stress within the vortex led to a greater scatter in fragment size distributions. In Taylor\textendash Couette systems, \citet{WANG2005} demonstrated that latex particle aggregates in NaCl solutions initially form intricate structures that subsequently restructure and fragment into more compact clusters. Likewise, \citet{GUERIN2017} showed that although hydrodynamic conditions determine aggregate size, the morphology ultimately reaches a steady state following cycles of lower and higher shear rates in turbulent Taylor\textendash Couette flow. These findings indicate that the increase in $\Tilde{F}$ in Fig.~\ref{fig:particle_images_clusters_SEM} may be attributed to the gradual elongation of the larger clusters caused by the hydrodynamic shear imposed on the solution by the movement of the cylinder. Once the shearing and breakup of the clusters proceed to form a higher fraction of individual spherical particles and more compact cluster fragments, $\Tilde{F}$ decreases. In addition, the individual particles in the clusters dissolve, leading to a decrease in their size. Therefore, the stable cluster fragments are again subject to restructuring as the reaction progresses.

Although the observed clustering of the particles may more accurately reflect their initial state during the dissolution process as a part of electrochemical reduction, this phenomenon complicates the estimation of the dissolution state. The effects of cluster breakup and dissolution were confounded using optical techniques, since both lead to a decrease in $\Tilde{d}$. To avoid particle clustering, the surfactant concentration was increased in preliminary tests, but this degraded the image quality due to the formation of streaks, which led to excess bubble formation and foaming induced by cylinder rotation. The clusters were observed in the dry powder, possibly due to tribological forces during transport or particle collection, so changing the surfactant conditions will not prevent these initial clusters. The use of a surfactant during the preparation of stock solutions would introduce an additional source of variability in the experiments and the interaction time between the particles and surfactant would have to be controlled. Furthermore, an ultrasonic cleaner could destroy some of the hollow structures of the particles. Thus, to separate the effects of cluster breakup and particle dissolution, the mean cluster breakup rate was estimated.

\subsection{Influence of temperature and light on the dissolution process}
\label{subsec:temp_and_light_influence}
\begin{figure}[ht!]
    \centering
        \begin{tikzpicture}
        	\node[inner sep=0pt] at (0,0)
        	{\includegraphics{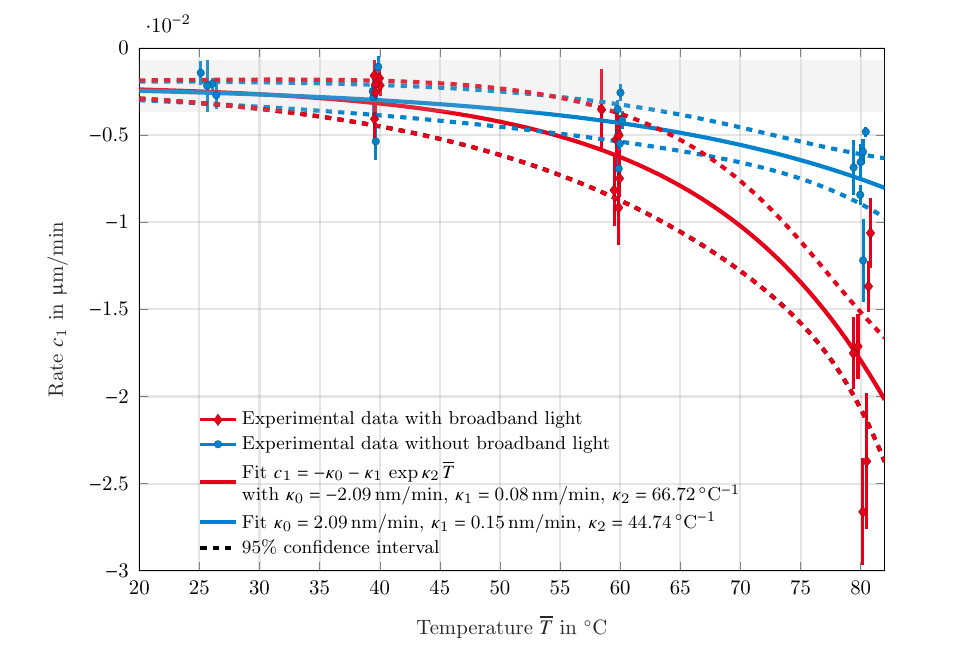}};
        \end{tikzpicture}
    \caption{Estimated $\mathrm{d}\Tilde{d}/\mathrm{d}t$ ($c_1$) as a function of temperature both with and without broadband light, fitted by an exponential function. The gray shaded marks the minimum and maximum $c_1$ CI value of fragmentation.}
    \label{fig:fit_rates}
\end{figure}

To estimate the influence of particle breakup on $c_1$, the temperature was lowered to room temperature (approximately 25$^\circ$C) for $\Phi=1.5$ without broadband light, whereas the other parameters specified in Tb.~\ref{tb:Factor_levels_reaction} were kept constant. The evolution of $\Tilde{d}(t)$ is shown in Fig.~\ref{fig:fragmentation_diameter_evolution} in the Appendix. This adjustment aimed to maintain the state of the governing interparticle forces responsible for the formation and breakup of clusters, including the pH, ionic strength, dielectric constant of the solvent, particle concentration, and hydrodynamic shear rate. Similarly, the reported low reaction rates at temperatures below 35$^\circ$C~\citep{VEHMAANPERAE2022} significantly reduced the influence of dissolution on $c_1$. As shown in Fig.~\ref{fig:fit_rates}, the rates obtained for fixed boundary conditions at $\overline{T}<40^\circ$C are statistically indistinguishable from those obtained at $\overline{T}\approx40^\circ$C. This indicates that, up to 40$^\circ$C, cluster breakup is still the dominant mechanism rather than particle dissolution. This finding is further supported by the increasing number of detected particles $n/n_\mathrm{max}$ up until the end of the experiment (see Fig.~\ref{fig:raw_data_all_n}). Cluster fragmentation contributes indirectly to the dissolution profile by increasing the total available surface area~\citep{LAUSCH2024}.

Interestingly, at 40$^\circ$C, light did not contribute significantly to $c_1$, and the influence of light became apparent only at higher temperatures. Consequently, the higher $c_1$ values at higher temperatures cannot only be caused by the higher activity of the surfactant, which is a function of temperature alone. The interaction between temperature and short\textendash wavelength light is a surprising result considering that \FetOt particles dissolve faster under the same light exposure at lower temperatures~\citep{LAUSCH2024}. In addition, the \FetOt particles used in the previous experiments were a combination of rhombohedral and spinel crystal structures~\citep{LAUSCH2024b}, similar to CIPs. Considering the schematic overview of reaction mechanisms in Fig.~\ref{fig:schematic_mechanisms}, in the present case the internal structure of the iron oxides is the decisive factor rather than the bulk crystal structure. As shown by \citet{CHOISEZ2022}, the particles were composed of a hematite shell surrounding a perforated core of magnetite interspersed with hematite spikes. At 40$^\circ$C, the reaction rate associated with the hematite shell may be so low that the concentration of dissolved iron in the solution is insufficient for the additional light\textendash induced reduction of some complexes, resulting in a negligible influence on the dissolution rate. As the temperature increases, so does the concentration of \Fethp complexes, which can be reduced by short\textendash wavelength light. Consequently, at 60$^\circ$C, a shift towards higher rates was observed ($c_1\approx-0.63\cdot10^{-2}\,\upmu$m/min), which is significantly different from the fragmentation rate of $c_1\approx-0.29\cdot10^{-2}\,\upmu$m/min. This trend continues at 80$^\circ$C, where the greatest difference is observed between the rates with and without broadband light. There is also a compounding effect for increased \Fethp concentration, since increasing concentrations also increase the \Fetwp quantum yield~\citep{WELLER2013} due to a secondary reduction of ferric complexes by the \ce{CO2} radicals. This secondary reduction pathway of the aqueous iron complexes is significant, as its contribution also explains the lower quantum yields observed under continuous irradiation compared to flash photolysis~\citep{WELLER2013}. Consequently, a higher concentration of ferric oxalato complexes results in an increased quantum yield, which elevates the production of ferrous ions. This, in turn, accelerates the reaction rate and leads to a higher overall iron ion concentration in solution.

\begin{figure}[ht!]
\centering
    \begin{minipage}[b]{0.43\textwidth}
    	\begin{tikzpicture}
    		\node[inner sep=0pt] at (0,0)
    		{\includegraphics{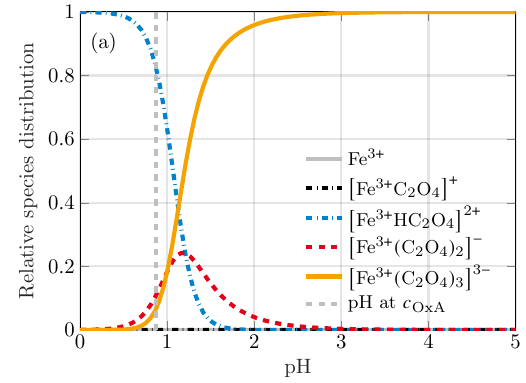}};
    	\end{tikzpicture}
    \end{minipage}    
    \hspace{1cm}
    \begin{minipage}[b]{0.47\textwidth}
    	\begin{tikzpicture}
    		\node[inner sep=0pt] at (0,0)
    		{\includegraphics{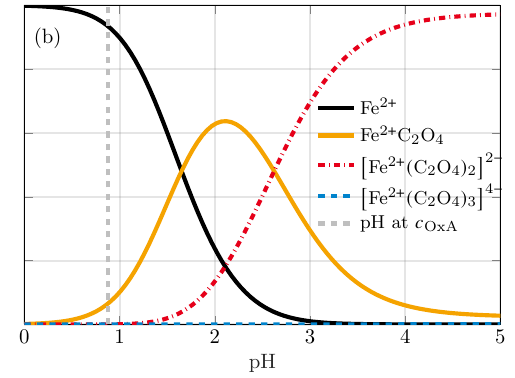}};
    	\end{tikzpicture}
    \end{minipage}
    \caption{Relative distribution of (a) ferric and (b) ferrous species as a function of pH, calculated using the equilibrium constant values from \citet{PANIAS1996B} and \citet{POZDNYAKOV2008}. Also shown is the pH of the $c_{\mathrm{H_2C_2O_4}}=0.45\,$M solution used in the present study, calculated assuming the dissociation of OxA with $\mathrm{pKa_1} = 1.25$ and $\mathrm{pKa_2} = 4.21$ at 25$^\circ$C~\citep{PANIAS1996B}.}
    \label{fig:speciation_actual_concentration}
\end{figure}

Both the OxA concentration and the pH affect the resulting distribution of iron oxalate complexes. In addition, increasing the temperature reduces the pH, as discussed previously~\citep{LAUSCH2024}. An overview of the ferric and ferrous complex distribution is provided in Fig.~\ref{fig:speciation_all_fe2_fe3} in the Appendix. From the rate data presented in Fig.~\ref{fig:fit_rates}, it can be concluded the presence of \Fetwp is beneficial for accelerating the dissolution process in the beginning of an experiment and is desirable in the context of iron oxide reduction via an electrochemical pathway. The addition of \Fetwp, either directly~\citep{BLESA1987,LITTER1992,TAXIARCHOU1997a,AMBIKADEVI2000,SUTER1988} or in the form of \Fetwpc-containing oxides~\citep{LEE2006}, to achieve a targeted increase in reaction rate is known from the literature.\\
For the OxA concentration used in this study, approximately 6\% of \Fetwp existed in the form of \ce{Fe^{II}C2O4} at equilibrium (Fig.~\ref{fig:speciation_actual_concentration}\,(b)). The figure shows that \ce{Fe^{II}C2O4} is the only stable \Fetwp complex in solution that can participate in the autocatalytic pathway (Eq.~\ref{equ:chem_reductive_dissolution_autocatalytic}), because all remaining ferrous ions are non\textendash complexed. 

If the pH is adjusted independently, as in the experiments conducted by \citet{LEE2007}, the maximum reaction rate coincides with the peak of \ce{Fe^{2+}C2O4} in equilibrium with the OxA concentration (Fig.~\ref{fig:speciation_Lee_concentration} in the Appendix) and decreases with increasing pH. Therefore, even if other ferrous complexes are present, \ce{Fe^{2+}C2O4} remains the dominant complex participating in the autocatalytic pathway. Thus, it is proposed that Eq.~\ref{equ:chem_reductive_dissolution_autocatalytic} can be reformulated to Eq.~\ref{equ:chem_reductive_dissolution_autocatalytic_specific} for the known oxidized complex:
\begin{subequations}
\begin{align}
\left[\rangle \mathrm{Fe}^{\mathrm{III}}\,\mathrm{Ox}^{\mathrm{n-}}\right]+\left[\mathrm{Fe}^{2+}\,\mathrm{C_2O_4}\right]_{(\mathrm{aq})} \quad & 
\rightarrow\quad 
\rangle \mathrm{Fe}^{\mathrm{III}}\,\mathrm{Ox}^{\mathrm{n-}} \ldots \mathrm{Fe}^{2+}\,\mathrm{C_2O_4}^\mathrm{2-} \\
\rangle \mathrm{Fe}^{\mathrm{III}}\,\mathrm{Ox}^{\mathrm{n-}} \ldots \mathrm{Fe}^{2+}\,\mathrm{C_2O_4}^\mathrm{2-} \quad & \rightarrow\quad \rangle \mathrm{Fe}^{\mathrm{II}}\,\mathrm{Ox}^{\mathrm{n-}} \ldots \mathrm{Fe}^{3+}\,\mathrm{C_2O_4}^{\mathrm{2-}} \\
\rangle \mathrm{Fe}^{\mathrm{II}}\,\mathrm{Ox}^{\mathrm{n-}} \ldots \mathrm{Fe}^{3+}\,\mathrm{C_2O_4}^{\mathrm{2-}} \quad & \rightarrow\quad \rangle \mathrm{Fe}^{\mathrm{II}}\,\mathrm{Ox}^{\mathrm{n-}}+\left[\mathrm{Fe}^{3+}\,\mathrm{C_2O_4}\right]^\mathrm{+}_{(\mathrm{aq})} \\
\rangle \mathrm{Fe}^{\mathrm{II}}\,\mathrm{Ox}^\mathrm{n-} \quad & \rightarrow\quad \left[\mathrm{Fe}^{2+}\,\mathrm{Ox}\right]^\mathrm{2-n}_{(\mathrm{aq})}\quad\quad\text{.}
\end{align}
\label{equ:chem_reductive_dissolution_autocatalytic_specific}
\end{subequations}
Given that the fraction of $\left[\mathrm{Fe}^{3+}\,\mathrm{C_2O_4}\right]^\mathrm{+}$ is only $\approx0.03$\% at $c_{\mathrm{OxA}}=0.45$\,M, this complex in Eq.~\ref{equ:chem_reductive_dissolution_autocatalytic_specific}c is transformed into more stable species, as shown in Fig.~\ref{fig:speciation_actual_concentration}\,(a). For the present case, approximately $82$\% of the ferric oxalate complexes in solution is $\left[\mathrm{Fe}^{3+}\,\mathrm{HC_2O_4}\right]^\mathrm{2+}$, which also suggest that photoreduction in the solution predominantly occurs via this complex. Note that information regarding the quantum yield for the reduction of \Fethp is limited. However, it is also only available in negligible amounts; therefore, even in the unlikely event of a high quantum yield (see \ref{sec:aqueous_iron_oxalate_complexes}), its contribution to the photochemical reduction process is assumed to be minor.

Although the presence of \ce{Fe^{2+}C2O4} accelerates the dissolution process, it has an adverse effect once its solubility limit is reached, leading to the precipitation of a solid product. When oxide particles act as nucleation points, a solid product layer forms around them, hindering further dissolution. In this case, the particle dissolution rate could be limited by diffusion of the reactants through that product layer rather than the reaction kinetics, as proposed by \citet{LEE2006}. Even in the case of homogeneous precipitation, iron is still lost to the solid phase, complicating subsequent processing steps in the context of electrowinning. As shown in Fig.~\ref{fig:speciation_actual_concentration}, the fraction of ferrous oxalate in the solution remains substantial at higher pH values because of the higher proportion of doubly deprotonated oxalate ions \ce{C2O4^{2-}}. To completely prevent the formation of ferrous oxalate, the pH can be lowered toward zero, at which point the fraction of non\textendash complexed ferrous iron approaches one. However, this would likely result in slower reaction rates because \ce{Fe^{2+}C2O4} appears to be the dominant complex driving the autocatalytic mechanism. To mitigate these effects, the concentration of OxA can be increased to its solubility limit (approximately 1\,M at room temperature~\citep{SANTAWAJA2021}). However, this does not address other potential influences such as the dynamic nature of the process, which renders the equilibrium diagrams shown in Fig.~\ref{fig:speciation_actual_concentration} invalid. Furthermore, the pH changes during the reaction depending on the initial OxA concentration and pH because the formation of oxalate\textendash containing complexes leads to the subsequent dissociation of OxA (Eq.~\ref{equ:chem_acid_dissociation}), as discussed in detail by \citet{SANTAWAJA2021}. There may also be local effects, as suggested by \citet{VEHMAANPERAE2022}, where the depletion of H$^+$ near the particle surface leads to a local pH that is higher than the bulk solution pH. Although it is possible to adjust the concentration of iron oxide in the solution such that the maximum ferrous oxalate concentration remains below its solubility limit, this approach is inefficient from an industrial perspective. Additionally, estimating the exact concentration of ferrous iron can be challenging owing to dynamic and interconnected boundary conditions. Another method for suppressing the formation of ferrous oxalate precipitate is the use of an OxA/nitric acid mixture, as described by \citet{VEHMAANPERAE2022}. Their study reports experiments with an excess of iron oxide to determine solubility limits, and the pH at the end of dissolution reached 0.6 for both magnetite and hematite. Thus, by combining these results with the species distributions shown in Fig.~\ref{fig:speciation_actual_concentration}\,(b), it appears that the role of nitric acid in preventing ferrous oxalate formation extends beyond lowering the pH because the reported pH values still fall within the stability range of \ce{Fe^{2+}C2O4}.

\begin{figure}[ht!]
    \centering
        \begin{tikzpicture}
        	\node[inner sep=0pt] at (0,0)
        	{\includegraphics{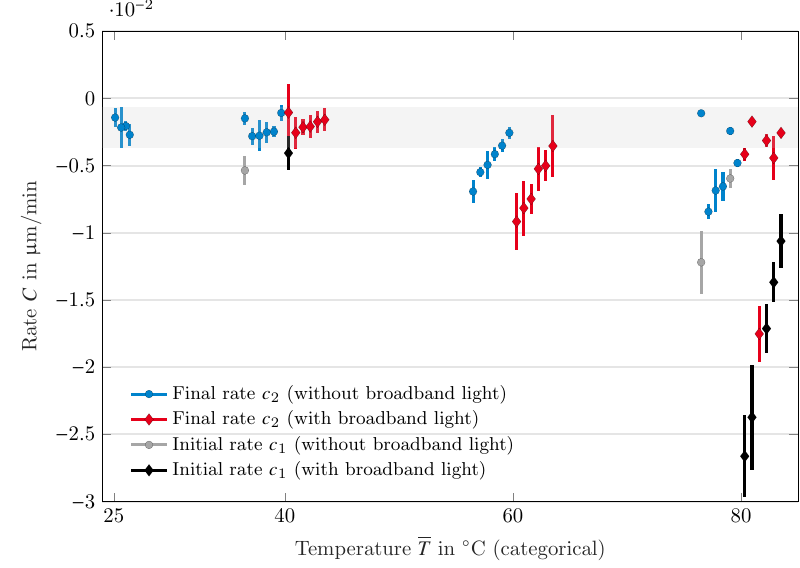}};
        \end{tikzpicture}
    \caption{Estimated $\mathrm{d}\Tilde{d}/\mathrm{d}t$ ($c_1$ and $c_2$) as a function of temperature both with and without broadband light irradiation. For better visibility, the data has been distributed and ordered according to the $c_1$ value at each temperature.}
    \label{fig:segmentation_rates}
\end{figure}

Further insights into the role of ferrous iron in the dissolution of CIPs are provided by the segmented experimental runs, which have been fitted using two linear models, as described in Sec.~\ref{subsec:determination_of_significant_factors}. The majority of segmented rates occurred at experiments at 80$^\circ$C, however there are also two experimental runs at lower temperatures, as shown in Fig.~\ref{fig:segmentation_rates}.

At the elevated temperature of 80$^\circ$C, almost all experiments with light exposure exhibit two significantly different rates, as also evident from the unfitted data in Fig.~\ref{fig:raw_data_all}. Given the complex species distribution and the internal iron oxide structure of the CIPs, it is likely that the formation of a solid product layer of ferrous oxalate products is responsible for the decrease in reaction rate.\\
In the present study, \Fetwp for this formation can originate from multiple sources; however, at the beginning of the dissolution experiments, the most significant influence is likely the short\textendash wavelength light exposure because the hematite shell of the CIPs contains no \Fetwpc. As described in Sec.~\ref{sec:Introduction}, short\textendash wavelength light can reduce terminal \Fetwpc in the hematite shell directly through a surface\textendash attached complex through a ligand\textendash to\textendash metal charge transfer, or reduce the stable ferric complexes. However, if ligand\textendash promoted reduction as described in Sec.~\ref{sec:Introduction} were the dominant interaction with short\textendash wavelength light, then the rates $c_1$ in Fig.~\ref{fig:segmentation_rates} should vary significantly at lower temperatures. In that scenario, the effect of short\textendash wavelength light would be directly tied to the surface area of the particles available for forming surface complexes. Particularly at $\overline{T}\approx40^\circ$C, clusters fragment at the same rate regardless of light irradiation, thereby providing a consistent available surface area. Consequently, if ligand\textendash promoted reduction significantly contributes to the \Fetwp concentration, a pronounced difference in $c_1$ should be observed under these conditions. Similarly, if the ligand\textendash promoted reduction of surface iron primarily led to the formation of a ferrous oxalate product layer, there would be a higher likelihood of segmented experimental runs for experiments with broadband light at temperatures of $\overline{T}\approx40^\circ$C and $\overline{T}\approx60^\circ$C. As evident from both the unfitted data in Fig.~\ref{fig:raw_data_all} and the fitted rates in Fig.~\ref{fig:segmentation_rates}, this is not observed. Thus, the influence of broadband light exposure appears to arise from the activation of Eq.~\ref{equ:chem_reductive_dissolution_autocatalytic_specific} through the reduction of aqueous ferric complexes, predominantly $\left[\mathrm{Fe}^{3+}\mathrm{HC_2O_4}\right]^\mathrm{2+}$.

As the reaction progresses and the magnetite cores of the particles are exposed to OxA, \Fetwpc within the particles becomes available, eliminating the need for \Fethp or \Fethpc reduction to form ferrous complexes. To form \ce{Fe^{2+}C2O4}, doubly deprotonated oxalate ions are required (Eq.~\ref{equ:chem_iron_oxalate_synthesis}). Although the fraction of \ce{C2O4^{2-}} in equilibrium diagrams is only $\approx0.01$\% at the oxalic acid concentration used (see e.g. \citep{LAUSCH2024}), the absolute concentration of $c_\mathrm{OxA}=0.45$\,M yields $c_{\mathrm{C_2O_4^{2-}}}$ in the order of $\mathcal{O}(10^{-5})$\,M. This is comparable to the concentrations employed in experiments with lower OxA concentration in the mM range, such as those by \citet{LITTER1991}, where solid ferrous precipitates formed under short\textendash wavelength light exposure. Consequently, \ce{C2O4^{2-}} can still react with the \Fetwpc ions of magnetite to form \ce{Fe^{2+}C2O4} directly at the particle surface. Magnetite dissolution was shown in a previous study~\citep{LAUSCH2024b} to occur via the reducing pathway. In that study, reaction products were continuously removed from the site, which suppressed the autocatalytic mechanism. The observations in that study indicate the formation of an amorphous iron\textendash oxalate structure. This in contrast to other studies~\citep{VEHMAANPERAE2022,VEHMAANPERAE2021A,SANTAWAJA2021} that reported the formation of crystalline iron oxalate dihydrate (humboldtine, $\left(\mathrm{Fe}^{\mathrm{II}} \mathrm{C}_2 \mathrm{O}_4 \cdot 2\,\mathrm{H}_2\mathrm{O}\right)$). This difference suggests that interactions with reaction products are essential for the formation of solid, crystalline ferrous oxalate.\\
The CIP dissolution observed by \citet{LAUSCH2024b} proceeded via a non\textendash reducing pathway\,\textendash\,similar to the hematite sample\,\textendash\,when the autocatalytic pathway was suppressed. In the present experiments, where the interaction with the reaction products per Eq.~\ref{equ:chem_reductive_dissolution_autocatalytic_specific} is permitted, Fig.~\ref{fig:fit_rates} indicates that the absence of a crystalline precipitate in ref.~\citep{LAUSCH2024b} may also result from the reaction not having progressed sufficiently to expose the magnetite core. Should the magnetite core be exposed at different times across particle sizes, one would expect a gradual transition from rate $c_1$ to $c_2$. However, based on the results of \citet{GOURSAT1973}, \citet{MI2022} state that the solid\textendash state growth rate of the irregular hematite layer depends mainly on the ambient oxygen concentration during the combustion process. Consequently, a uniform hematite shell thickness is expected across all particle sizes when curvature effects are neglected. Under surface\textendash reaction\textendash controlled dissolution, the particle radius decreases at a rate independent of its initial size, causing most particles to expose their magnetite cores nearly simultaneously. Electron\textendash backscatter diffraction data from \citet{CHOISEZ2022} (cf. Fig. 9) was analyzed to estimate that the shell thickness of combusted iron particles ranges between 2.0 and 2.5\,$\upmu$m. This finding aligns with experimental data that smaller particles exhibit higher degrees of oxidation~\citep{BUCHHEISER2023} because a constant shell thickness would form a larger proportion of their total mass. Given that most particles are approximately 8–10\,$\upmu$m in diameter~\citep{BUCHHEISER2023}, even variations in hematite shell thickness across particle size are expected to have only a limited impact on the change from $c_1$ to $c_2$ because the magnetite core in the majority of particles becomes exposed nearly simultaneously through dissolution. Assuming that fragmentation is not temperature\textendash dependent, one can estimate the time most of the shell will have dissolved under 80$^\circ$C broadband light exposure—the condition observed in most segmented experimental runs. By subtracting the fragmentation rate from the shrinkage rate, an effective dissolution rate of approximately $1.51\cdot10^{-2}\,\upmu$m/min is obtained. Consequently, the shell should be completely dissolved between approximately 265 min and 332 min for initial shell thicknesses of 2.0 and 2.5\,$\upmu$m, respectively (i.e., particle diameters of 4.0\,$\upmu$m and 5\,$\upmu$m). However, as shown in Fig.~\ref{fig:raw_data_all}, these predicted times far exceed the experimentally observed transition in $C$ ($123\pm31$ min), despite significant scatter in the data (earliest onset  $\approx82$ min; latest $\approx148$ min). This discrepancy may reflect an overestimation of shell thickness, inhomogeneous shell morphology, or a dominant contribution from homogeneous precipitation in solution instead of exposure of the magnetite in the particles' core. Moreover, the abrupt decrease in reaction rate is likely driven by overlapping processes once a substantial fraction of the particles has dissolved: light‑assisted reduction of ferric iron–oxalate complexes promoting ferrous oxalate precipitation, exposure of the magnetite core to OxA once the shell is dissolved, or a combination of both.

\section{Conclusion}
This study examined the effects of temperature (40-80$^\circ$C) and broadband light exposure (190-2500\,nm) on the dissolution rate of CIPs at three different fuel\textendash to\textendash air equivalence ratios ($\Phi$: 0.67, 1.0, and 1.5) in OxA (0.45\,M).
The optical analysis revealed that the particles existed as clusters, which fragmented during dissolution due to interactions with the fluid flow, other clusters, and walls of the reaction vessel. The combined effects of cluster breakup and dissolution resulted in an almost linear decrease in median particle diameter over time. 
Statistical analysis revealed that both broadband light exposure and increased temperature enhance the particle dissolution rate. A temperature above 40$^\circ$C is required to achieve a particle shrinkage rate that significantly differs from the baseline cluster fragmentation rate, leading to a diameter decrease of approximately $0.29\cdot10^{-2}\,\upmu$m/min. Furthermore, the effect of broadband light on dissolution intensifies with rising temperature. At 80$^\circ$C under broadband light irradiation, a shrinkage rate of approximately $1.8\cdot10^{-2}\,\upmu$m/min can be achieved, albeit with considerable data scatter. This temperature\textendash dependent influence of broadband light is attributed to the slow light\textendash and temperature\textendash dependent dissolution reaction of the hematite shell of the CIPs. Once a sufficient concentration of ferric oxalate complexes is available in the solution for effective photo\textendash reduction, there is a subsequent acceleration of dissolution through activation of an autocatalytic mechanism. Because the effect of broadband light exposure is negligible at 40$^\circ$C, ligand\textendash promoted reduction at the particle surface appears to play only a minor role in the overall dissolution process.
Although the literature suggests that increasing $\Phi$ enhances the magnetite fraction, statistical analysis revealed that $\Phi$ does not significantly affect the dissolution rate. This is surprising considering that magnetite increases the reaction rate through both its spinel\textendash type crystal structure and ferrous ions that activate the autocatalytic pathway. The observed lack of influence of $\Phi$ may be because the effect is too small to be detected using the current method or could be due to hematite interstitials blocking access to the magnetite core, regardless of the fraction of magnetite in the particle.
At 80$^\circ$C, particularly under light exposure, a sudden decrease in the particle shrinkage rate was observed during dissolution. This may be a result of the homogeneous formation of solid ferrous oxalate once the solubility limit was reached, with the remaining particles serving as nucleation sites for the solid precipitate. Additionally, a solid product layer may form owing to the exposure of the magnetite core, leading to a sudden increase in available ferrous iron directly on the particle surface. These two mechanisms are not mutually exclusive and may simultaneously contribute to the observed behavior. The formation of a solid precipitate is undesirable in electrowinning processes because it reduces the reaction rate and results in the conversion of iron to a product requiring further processing. \\
Considering these results and the stable iron\textendash oxalate complexes, it is concluded that the efficient dissolution of CIPs requires a temperature above 60\,$^\circ$C and the maximum concentration of OxA, which has a solubility limit of approximately 1\,M at room temperature. If independent pH adjustment is feasible, lower values are preferable to limit the formation of ferrous oxalate below its solubility threshold. However, the complete suppression of ferrous oxalate is not advisable, as this would significantly slow down dissolution because of the need for ferrous iron to activate the autocatalytic mechanism that increases the dissolution rate via a self\textendash reinforcing process. Considering the structure of the studied CIPs, this initial activation may require short\textendash wavelength light or the addition of ferrous salts because the hematite shell obstructs access to the magnetite core.

\section*{Author contributions}
\textbf{M. Lausch}: Writing – original draft (lead), Visualization (lead), Validation (lead), Software (lead), Methodology (equal), Investigation (equal), Formal analysis (lead), Conceptualization (supporting). 
\textbf{Y. Ruan}: Investigation (equal). 
\textbf{P. Brockmann}: Writing – original draft (supporting), Writing – review \& editing (equal).
\textbf{A. Zimina}: Writing – original draft (supporting), Writing – review \& editing (equal). 
\textbf{B.J.M. Etzold}: Writing – review \& editing (equal), Supervision, Resources, Methodology (equal), Conceptualization (equal)
\textbf{J. Hussong}: Writing – review \& editing (equal), Supervision, Resources, Funding acquisition (lead), Conceptualization (equal).

\section*{Acknowledgements}
This work was funded by the Hessian Ministry of Higher Education, Research, Science and the Arts - cluster project Clean Circles. Thanks to Reda Kamal for his insightful discussions. We thank the Institute of Materials Science at the Technical University Darmstadt for the access to the scanning electron microscope and J. Schmidpeter and P. Neuhäusel in particular for their help during its operation. We thank Arne Scholtissek for his valuable input and constructive suggestions that have enhanced this manuscript. Parts of this work have been improved grammatically and stylistically using language models, including ChatGPT (OpenAI) and DeepL Write (DeepL SE).

\bibliography{bib_dissolution.bib}

\appendix

\section{Iron electrowinning}
\label{sec:Iron_electrowinning}
Alkaline slurry electrolysis has emerged as the dominant research pathway, with an European Union initiative starting in 2004 with the ULCOWIN project~\citep[p.32]{EUULCOWIN}. In that process, alkaline suspensions containing pure iron oxides or residues from industrial processes are used. Alternative approaches are less frequently explored, such as using iron oxide pellets as cathodes~\citep{HAARBERG2020a}. In pursuit of decarbonizing ironmaking, several European Union projects aimed at enhancing the technological maturity of iron EW through projects like IERO~\citep{EUIERO2016} and, more recently, Siderwin~\citep{EUSIDERWIN}. The targeted market entry for this technology has been pushed back from 2040~\citep[p.20]{EUSCEN2015} up to 2045~\citep[p.26]{EUTCH2022}. Outside Europe, the U.S. ARPA‑E ROSIE program~\citep{ARPAE_ROSIE} has very recently launched initiatives to optimize iron EW~\citep{ARPAE_FERRIC}.\\
Generally, most research has focused on alkaline electrolytes, typically at temperatures around 100\,$^\circ$C. Studies by \citet{ALLANORE2007,ALLANORE2008,ALLANORE2010a,ALLANORE2010b} in a 50 wt\% NaOH--\ce{H2O} electrolyte reveal that direct solid\textendash state reduction occurring during particle collisions with the cathode is the dominant pathway. This mechanism yields current efficiencies up to 98\% for hematite, whereas iron hydroxide and magnetite achieve much lower efficiencies (as low as 5\% for magnetite \citep{FEYNEROL2017}; see also \citep{HAARBERG2020b}). Moreover, if solid\textendash state reduction is inactive, the low solubility of iron in \ce{NaOH} (approximately $2\cdot10^{-3}$ M, as cited in \citet{ALLANORE2007} from \citet{PICARD1980}) can impair efficiency. This solubility decreases further at lower temperatures or \ce{NaOH} concentrations~\citep{ISHIKAWA1997}. Thus, novel electrolytic cell designs are needed to exploit the solid\textendash state reduction effect (see e.g. \citep[p.70]{EUIERO2016}).\\
In contrast, acidic electrolytes enable dissolution without solid\textendash state reduction but face two challenges: poor deposit quality (brittle, felt\textendash like deposits) and reduced current efficiency~\citep{MAJID2023}. The lower efficiency arises primarily from the dominant hydrogen evolution reaction (HER) at the cathode and the cycling of \Fetwp and \Fethp ions due to the coexistence of ferrous and ferric species~\citep{LOPES2024}. Comparative studies by \citet{MAJID2023} showed that concentrated \ce{NaOH} outperforms hydrochloric acid (\ce{HCl}), yielding higher current efficiencies and more manageable deposits with fewer hematite particles. Although \ce{HCl} enhances dissolution through \ce{Fe-Cl} complex formation~\citep{SIDHU1981}\citep[p.301]{CORNELL2003}, its performance leaves room for improvement. As stated in Sec.~\ref{sec:Introduction}, OxA is a promising option due to its high dissolution rates, reducing power, and iron oxide solubility limit. \citet{LOPES2024} concluded that the rapid dissolution and complexing ability of the OxA are key for enhancing the deposition step. However, they observed the formation of ferrous oxalate products at the iron and platinum cathode in sulfuric and oxalic acid mixtures at 23$^\circ$C and 50$^\circ$C. They also reported brittle, dendritic deposits and a high HER contribution, resulting in current efficiencies capped at 36\%. \citet{PAL2010} observed ferrous oxalate formation at a lead anode and improved efficiency when leaching at 90$^\circ$C followed by EW of the leachate at room temperature using a lead anode and steel cathode. Importantly, they employed a polymer diaphragm to compartmentalize the cell to reduce iron re\textendash oxidation due to the oxygen evolution reaction at the anode. Their experiments indicated that higher OxA concentrations increase current efficiency (up to approximately 19\% at 0.33\,M).\\
In summary, only a limited number of studies have explored the EW of iron in acidic media. When utilizing OxA, it appears that higher OxA concentrations and elevated temperatures improve current efficiencies and enhance the dissolution rate (see also Sec.~\ref{sec:Introduction}). Additionally, the precipitation of solid iron oxalate products is to be avoided by a suitable choice of boundary conditions.

\section{Occurring reaction mechanisms}
\label{sec:appendix_mechanisms_equations}

Diprotic OxA dissociates in two steps: 
\begin{equation}
\begin{aligned}
 \mathrm{H}_2 \mathrm{C}_2 \mathrm{O}_{4(\mathrm{aq})} & \leftrightarrow \quad \mathrm{HC}_2 \mathrm{O}_4{ }^{-}{ }_{\text {(aq) }}+\mathrm{H}^{+}{ }_{\text{(aq)}}\\
 \mathrm{HC}_2 \mathrm{O}_4{ }^{-}_\text {(aq) } & \leftrightarrow \quad \mathrm{C}_2 \mathrm{O}_4{ }^{2-}{ }_{(\mathrm{aq})}+\mathrm{H}^{+}{ }_{\text {(aq) }}
\end{aligned}    
\label{equ:chem_acid_dissociation}
\end{equation}

The first step in the dissolution process involves the protonation of a surface Lewis base oxide, resulting in a positively charged surface. This protonated surface serves as the starting point for the formation of a complex with an oxalate ion from solution \citep{PANIAS1996A,LITTER1992}. 

\begin{align}
\rangle \mathrm{Fe}^{\mathrm{III}}-\mathrm{O}+\mathrm{H}^{+} \,&\leftrightarrow \quad \rangle \mathrm{Fe}^{\mathrm{III}}-\mathrm{O} \ldots \mathrm{H}^{+}\label{equ:chem_surface_protonisation}\\
\rangle \mathrm{Fe}^{\mathrm{III}}-\mathrm{OH}^{+}+\mathrm{Ox}^{\mathrm{n}-}+\mathrm{H}^{+} & \leftrightarrow \left[\rangle \mathrm{Fe}^{\mathrm{III}}\,\mathrm{Ox}\right]^{2-\mathrm{n}}+\mathrm{H}_2 \mathrm{O}
\label{equ:chem_surface_protonisation_complex}
\end{align}

An additional non\textendash reducing dissolution pathway involves the iron oxalate surface complex, which reacts with an additional proton and desorbs into solution, as shown in Eq.~\ref{equ:chem_nonreductive_dissolution}~\citep{LITTER1992,PANIAS1996A}:

\begin{equation}
    \begin{aligned}
    \left[\rangle \mathrm{Fe}^{\mathrm{III}}\,\mathrm{Ox}^{\mathrm{n}-}\right]+\mathrm{H}^{+} \rightarrow\left[\mathrm{Fe}^{3+}\,\mathrm{Ox}\right]_{\,(\text{aq})}^{3-\mathrm{n}} + \,\,\rangle \mathrm{H}
    \label{equ:chem_nonreductive_dissolution}
    \end{aligned}
\end{equation}

For the reducing pathway, an electron transfer occurs as shown in Eq.~\ref{equ:chem_reductive_dissolution_induction}\,(a). If the doubly deprotonated oxalate forms the surface complex at low concentrations of OxA, the subsequent desorption process can be described as shown in Eq.~\ref{equ:chem_reductive_dissolution_induction}\,(b) \citep{PANIAS1996A}:

\begin{subequations}
\begin{align}
    \left[\rangle\mathrm{Fe}^{\mathrm{III}}\,\mathrm{Ox}^{\mathrm{n}-}\right] & \leftrightarrow \left[\rangle \mathrm{Fe}^{\mathrm{II}}\,\mathrm{Ox}^{-\mathrm{n}+1}\right]\label{equ:chem_electron_transfer}\\
    2\left[\rangle \mathrm{Fe}^{\mathrm{II}}-\mathrm{C}_2 \mathrm{O}_4^{-}\right]+2\,\mathrm{H}^{+} & \rightarrow 2\,\mathrm{Fe}_{(\mathrm{aq})}^{2+}+2\,\mathrm{CO}_2+\mathrm{C}_2 \mathrm{O}_4^{2-}+2\,\,\rangle \mathrm{H}\label{equ:chem_reductive_dissolution_induction}
\end{align}
\end{subequations}

Light\textendash driven homogeneous photolysis of the tris\textendash oxalato ferrate complex in solution is described in Eq.~\ref{equ:chem_photolysis_iron_oxalate_complex}~\citep{POZDNYAKOV2008}. However, as summarized in Sec.~\ref{sec:Introduction} and \ref{sec:interaction_of_light_and_iron_oxides}, other photoactive complexes may exist in solution. Note that formed \ce{CO2} radicals may react with other unphotolyzed ferric complexes.
\begin{equation}
2\left[\mathrm{Fe}^{\mathrm{3+}}(\mathrm{\mathrm{C}_2 \mathrm{O}_4})_3\right]^{3-} \xrightarrow[]{E_{\mathrm{photon}}} 2\left[\mathrm{Fe}^{\mathrm{2+}}(\mathrm{\mathrm{C}_2 \mathrm{O}_4})_2\right]^{2-}+2 \mathrm{CO}_2+\mathrm{\mathrm{C}_2 \mathrm{O}_4}^{2-}\,\,\text{.}
\label{equ:chem_photolysis_iron_oxalate_complex}
\end{equation}

The $\mathrm{pH}$, OxA concentration and presence of ferrous ions controls the formation of a solid precipitate, iron(II) oxalate dihydrate~\citep{VEHMAANPERAE2022}, once its concentration surpasses its solubility limit ($c_{\mathrm{FeC_2O_4}}\approx 447.2\,\upmu$M~\citep[p.487]{HOGNES1940} in water at room temperature):

\begin{equation}
\mathrm{Fe}^{\mathrm{2+}}+\mathrm{C}_2 \mathrm{O}_4^{2-}+2\,\mathrm{H}_2 \mathrm{O} \leftrightarrow \left(\mathrm{Fe}^{\mathrm{II}} \mathrm{C}_2 \mathrm{O}_4 \cdot 2\,\mathrm{H}_2 \mathrm{O}\right)_{(\mathrm{s})}\,\text{,}
\label{equ:chem_iron_oxalate_synthesis}
\end{equation}

Depending on the presence of dissolved oxygen in the solution, the ferrous ions in solution can be re\textendash oxidized~\citep{TAXIARCHOU1997b}:

\begin{equation}
    4\left[\mathrm{Fe}^{2+}\left(\mathrm{C}_2 \mathrm{O}_4\right)_2\right]^{2-}+\mathrm{O}_{2(\mathrm{aq})}+4\,\mathrm{H}^{+}+4\,\mathrm{C}_2 \mathrm{O}_{4}^{2-} \leftrightarrow 4\left[\mathrm{Fe}^{3+}\left(\mathrm{C}_2 \mathrm{O}_4\right)_3\right]^{3-}+2\,\mathrm{H}_2 \mathrm{O}\,\text{.}
    \label{equ:chem_ferrous_sol_re-oxidation}
\end{equation}

\section{Interaction of light and iron oxides}
\label{sec:interaction_of_light_and_iron_oxides}
\subsection{Aqueous iron oxalate complexes}
\label{sec:aqueous_iron_oxalate_complexes}
The photoreduction of the tris–oxalato ferrate(III) complex, $\mathrm{\left[Fe^{3+}(C_2O_4)_3\right]^{3-}}$, has been used in actinometers for nearly 70 years~\citep{HATCHARD1956} and is well documented in the literature. Its $\epsilon(\lambda)$ decreases from about $\mathcal{O}(10^4)$ mol$^{-1}$ cm$^{-1}$ in the UV to $\mathcal{O}(10^{-1})$ mol$^{-1}$ cm$^{-1}$ at 500 nm. As shown in Fig.~\ref{fig:light_overview}, fewer oxalate ligands yield lower UV absorption~\citep{POZDNYAKOV2008,WELLER2013}, though at 436 nm the bis–oxalato complex ($\approx62$ mol$^{-1}$ cm$^{-1}$) absorbs more than the tris–oxalato complex ($\approx24$ mol$^{-1}$ cm$^{-1}$)~\citep{FAUST1993}. Reported values of $\varphi(\lambda)$ vary in the literature. Despite some discrepancies (e.g. $\varphi(220,\mathrm{nm})=1.57\pm0.02$ \citep{GOLDSTEIN2008} vs. $\varphi(222,\mathrm{nm})=0.5$ \citep{FERNANDEZ1979}), it is generally accepted that for $\mathrm{\left[Fe^{3+}(C_2O_4)_3\right]^{3-}}$, $\varphi>1$ for $\lambda<$400 nm (e.g. 1.25 at 300 nm~\citep{HATCHARD1956}, 1.27 at 308 nm~\citep{WELLER2013}, 1.24 at 313 nm~\citep{GOLDSTEIN2008}). This is due to the formation of two \Fetwp ions per photon when sufficient unphotolyzed complexes enable secondary reductions via \ce{CO2} radicals ($\mathrm{CO_2^{- \cdot}}$) \citep{WELLER2013} (see also Eq.~\ref{equ:chem_photolysis_iron_oxalate_complex}). Consequently, the quantum yield is concentration\textendash dependent, which may account for discrepancies in reported values. Moreover, $\varphi(\lambda)$ decreases rapidly with wavelength, nearing zero between 580 and 590~nm \citep{STRAUB2018}; at $\lambda=436$ nm, values of $\varphi=1\pm 0.25$ for $\mathrm{\left[Fe^{3+}(C_2O_4)_2\right]^-}$ and $\varphi=0.6\pm0.46$ for $\mathrm{\left[Fe^{3+}(C_2O_4)_3\right]^{3-}}$ have been reported~\citep{FAUST1993}. Dissolved oxygen can reduce these yields~\citep{PARKER1959, DEMAS1981} up to tenfold in air\textendash saturated solutions~\citep{FAUST1993}, particularly at $\text{pH}<4$ due to hydrogen peroxide formation~\citep{ZUO1992}. However, as shown by \citet{JEONG2004}, the effect of oxygen also depends on the concentration of \Fethp in solution. At \Fethp concentrations of $10^{-3}$ M the rate of \Fetwp generation in oxygen\textendash saturated solution was nearly indistinguishable from that of nitrogen\textendash saturated solution due to the higher contribution of the reaction involving $\mathrm{CO_2^{- \cdot}}$ and $\mathrm{\left[Fe^{3+}(C_2O_4)_3\right]^{3-}}$. Data for mono–oxalato complexes remain limited; while \citet{COOPER1972} stated that mono\textendash , bis\textendash , and tris–oxalato complexes have similar quantum yields, they also assumed that photoaquation plays a dominant role during photoexcitation~\citep{DEGRAFF1971}. However, \citet{POZDNYAKOV2008} reported that photoaquation does not occur. Similarly, \citet{LONGETTI} investigated the photoexcitation mechanisms of ferrioxalate in detail and found that after a ligand\textendash to\textendash metal electron charge transfer there is a detachment of \ce{CO2} and \ce{CO2} radicals that triggers secondary reduction, with no indication of photoaquation mechanisms. Beyond these complexes, information on uncomplexed \Fethp is limited. In aqueous solution, hexaaquairon(III) is the predominant species at low total \Fethp concentrations ($\leq 5$ mM) and under acidic conditions ($\text{pH}<3$), where $\mathrm{[Fe^{3+}(H_2O)_6]^{3+}}$ and $\mathrm{[Fe(OH)(H_2O)_5]^{2+}}$ are the dominant iron(III) aqua ions~\citep{DANFORTH2017}. These species have molar absorption coefficients about one order of magnitude lower (200–350 nm in aqueous perchloric acid~\citep{TURNER1957}) than the ferric oxalate complexes and quantum yields on the order of $\mathcal{O}(10^{-3})$~\citep{FAUST1990,LANGFORD1975}. Therefore, the photo\textendash induced reduction of uncomplexed \Fethp is assumed to contribute insignificantly to the \Fetwp concentration.

\subsection{Direct interaction}
\label{subsec:direct_interaction}
In hematite the energy required for an \Fetwp, $\mathrm{Fe^{4+}}$ pair is approximately 2 eV \citep{GOODENOUGH1971}, with a value of 2.2 eV commonly reported in the literature~\citep{SHERMAN2005,XU2000}. This corresponds to a wavelength of $\lambda\lesssim564$ nm required to reduce \Fethpc in hematite. This value increases for decreasing pH and is relatively unaffected by temperature except for the indirect influence of temperature on pH~\citep{XU2000}. In contrast to the hematite semiconductor, magnetite is metallic at room temperature~\citep{GOODENOUGH1971}. The small band gap (approximately 0.1 eV~\citep{XU2000}) leads to a broad absorption across the light spectrum, explaining the black color of magnetite~\citep{GOODENOUGH1971}. Depending on the band gap of the iron oxide, the reducing dissolution of iron can occur if the half\textendash cell potential of this dissolution reaction is higher than the potential of the conduction band~\citep{SHERMAN2005}. For smaller particles in the nanometer scale, the band gap further increases with decreasing size~\citep{COLTON2014}. For the CIPs, there can be additional effects on the band gap due to the contact between metallic iron and magnetite, and semiconducting hematite such as band bending~\citep{ZHANG2012}. Band bending can also occur at the solid/liquid interface depending on the solution pH~\citep{STUMM1992}. For pure hematite, the direct photoreduction is improbable despite its semiconducting properties, attributed to a fast recombination of holes and electrons from the valence and conduction band \citep{LV2022}. Similarly, pure magnetite is also not directly reduced by light~\citep{BEYDOUN2002}.

\section{Statistical analysis}
\label{sec:appendix_statistical_analysis}
As stated in Sec.~\ref{subsec:determination_of_significant_factors}, ANOVA requires mutual independence, normal distribution, and equal variances of the error variable for the results to be valid. The test for mutual independence was performed by plotting standardized residuals versus run order, normal distribution was tested using a normal probability plot, and equal variances were tested by inspecting standardized residuals versus their absolute fitted value from ANOVA. As visible in Fig.~\ref{fig:fit_rates}, the $c_1$ variance increases with increasing temperature and is a function of light irradiation (heteroscedasticity). This was also confirmed by the significant fanning of non\textendash normally distributed standardized residuals (see Fig.~\ref{fig:residuals_normality_raw} and~\ref{fig:residuals_fitvals_raw}). This invalidates the inferences drawn from the ANOVA results~\citep{ERCEG2008}, although the robustness of the ANOVA against the violation of the prerequisites is debated~\citep{SCHMIDER2010,ITO1980}, and alternative criteria for the requirements have been explored~\citep{KIM2018}. Data were transformed to account for non\textendash normality and heteroscedasticity. Numerous methods for data transformation have been developed (see \citep[pp. 317-329]{BOX2005}, \citep[pp. 316-331]{COOK2009}, and \citep{YEO2000}) with the general goal of stabilizing variance and restoring a normal distribution. As the data do not include $\left|c_1\right|=0$, the widely used Box\textendash Cox transformation technique~\citep{BOX1964} is feasible. This transformation is defined as~\citep{BOX1964,ATKINSON2021}:
\begin{equation}
    \Xi =
    \begin{cases}
      \left(\left|c_1\right|^\zeta -1\right)/\zeta & \quad \zeta \neq 0\\
      \mathrm{ln}\left(\left|c_1\right|\right)     & \quad \zeta = 0\text{,}
    \end{cases}
    \label{equ:BoxCox_transformation}
\end{equation}
where the optimal exponent $\zeta=-0.048$ is the maximum\textendash likelihood estimate of the profile log\textendash likelihood obtained using MATLAB\textendash based statistical software~\citep{RIANI2012}. Owing to the requirement of positive data, the absolute value of $c_1$ was used. Because all $c_1$ values are smaller than zero and none of the values are exactly zero, this does not influence the validity of the interpretation. After inspecting the standardized residuals, ANOVA was computed again using $\Xi$. 

\begin{figure}[ht!]
\centering  
    \begin{minipage}[b]{0.45\textwidth}
    	\begin{tikzpicture}
    		\node[inner sep=0pt] at (0,0)
    		{\includegraphics{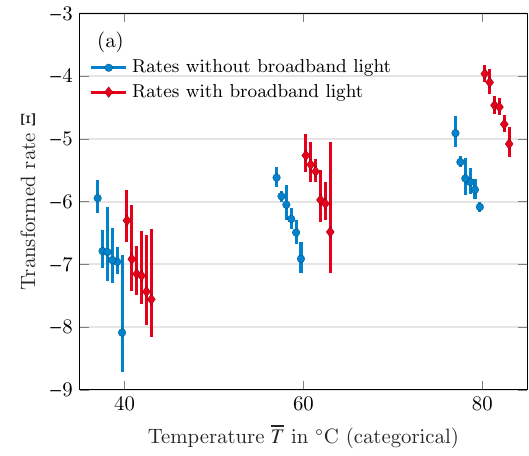}};
    	\end{tikzpicture}
    \end{minipage}    
    \hspace{1cm}
    \begin{minipage}[b]{0.45\textwidth}
    	\begin{tikzpicture}
    		\node[inner sep=0pt] at (0,0)
    		{\includegraphics{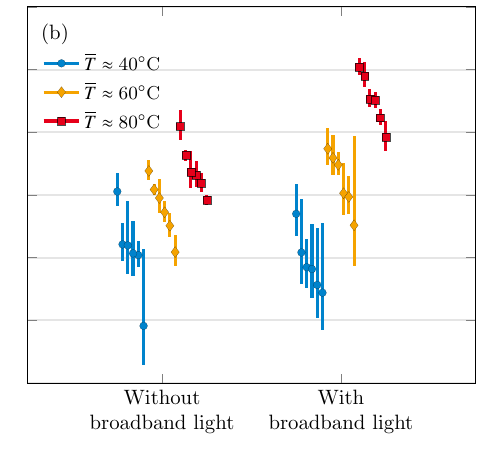}};
    	\end{tikzpicture}
    \end{minipage}
    \caption{(a) Effect of temperature on $\Xi$. For better visibility, the data has been distributed and ordered according to the value of $\Xi$ at each temperature. The measured values of $\overline{T}$ are shown e.g., in Fig.~\ref{fig:fit_rates}. The effect of broadband light at a given temperature can be inferred from the color coding. (b) Effect of broadband light on $\Xi$, color\textendash coded for each temperature.}
    \label{fig:main_effects_temp_light}
\end{figure}

\begin{figure}[ht!]
\centering
    \begin{minipage}[b]{0.45\textwidth}
        \begin{tikzpicture}
        	\node[inner sep=0pt] at (0,0)
        	{\includegraphics{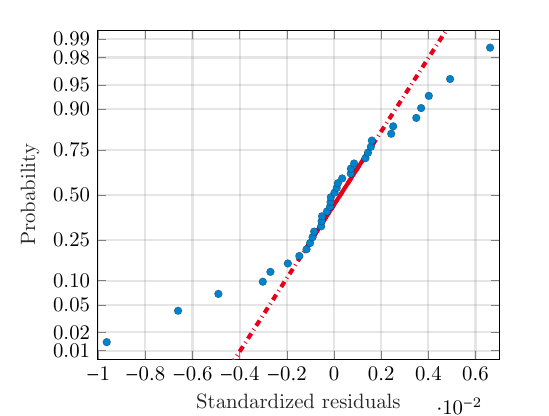}};
        \end{tikzpicture}
        \caption{Normal probability plot of untransformed $c_1$ data.}
        \label{fig:residuals_normality_raw}
    \end{minipage}    
    \hspace{0.8cm}
    \begin{minipage}[b]{0.45\textwidth}
        \begin{tikzpicture}
        	\node[inner sep=0pt] at (0,0)
        	{\includegraphics{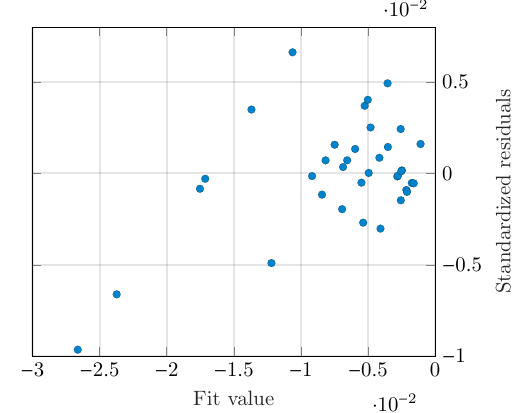}};
        \end{tikzpicture}
        \caption{Heteroscedasticity of untransformed $c_1$ data (ANOVA).}
        \label{fig:residuals_fitvals_raw}
    \end{minipage}
\end{figure}

To investigate the direction and magnitude of the change, the values of $\Xi$ are presented as functions of the significant terms in Fig.~\ref{fig:main_effects_temp_light}. The error bars indicate the transformed uncertainties of $c_1$ using Eq.~\ref{equ:BoxCox_transformation}. Note that for ANOVA calculations, the factor levels of the rates were decisive, as indicated by their respective markers. Focusing first on Fig.~\ref{fig:main_effects_temp_light}\,(a), increasing the temperature generally results in an increasing rate, which further increases under light irradiation. The effect of light was mainly visible at $\overline{T}\approx80^\circ$C, as shown in Fig.~\ref{fig:main_effects_temp_light}\,(a) and (b). An increase in the reaction rate in the presence of light was expected, given its influence on some of the mechanisms (Fig.~\ref{fig:schematic_mechanisms}).

\begin{figure}[ht!]
\centering
    \begin{minipage}[b]{0.45\textwidth}
    	\begin{tikzpicture}
    		\node[inner sep=0pt] at (0,0)
    		{\includegraphics{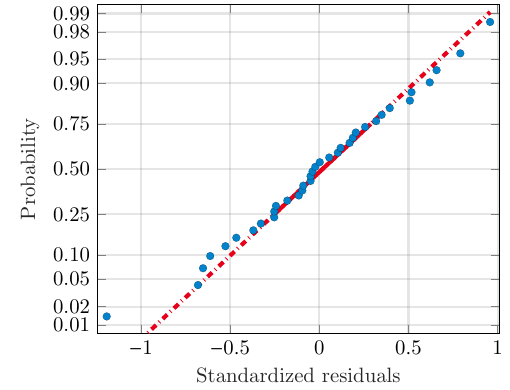}};
    	\end{tikzpicture}
        \caption{Normal probability plot of $\Xi$ data.}
        \label{fig:residuals_normality}
    \end{minipage}    
    \hspace{0.8cm}
    \begin{minipage}[b]{0.45\textwidth}
        \begin{tikzpicture}
        	\node[inner sep=0pt] at (0,0)
        	{\includegraphics{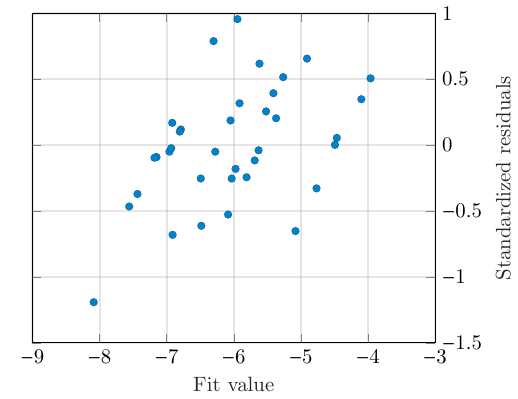}};
        \end{tikzpicture}
        \caption{Homoscedasticity of $\Xi$ data (ANOVA).}
        \label{fig:residuals_fitvals}
    \end{minipage}
\end{figure}

\begin{figure}[ht!]
\centering
    \begin{minipage}[b]{0.45\textwidth}
        \begin{tikzpicture}
        	\node[inner sep=0pt] at (0,0)
        	{\includegraphics{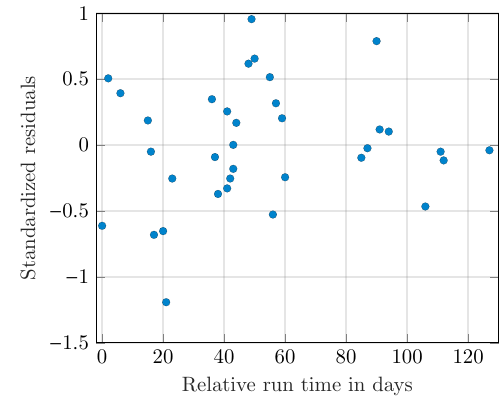}};
        \end{tikzpicture}
        \caption{Mutual independence plot of $\Xi$ data (ANOVA).}
        \label{fig:residuals_runtime}
    \end{minipage}    
    \hspace{0.8cm}
    \begin{minipage}[b]{0.45\textwidth}
        \begin{tikzpicture}
        	\node[inner sep=0pt] at (0,0)
        	{\includegraphics{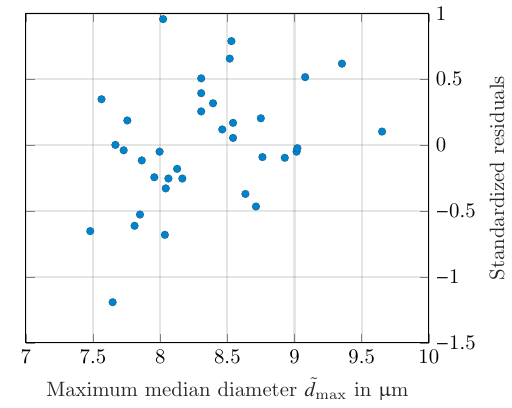}};
        \end{tikzpicture}
        \caption{Correlation plot of maximum diameter $\Tilde{d}_{\mathrm{max}}$.}
        \label{fig:residuals_diam}
    \end{minipage}
\end{figure}

\begin{figure}[ht!]
\centering
    \begin{minipage}[b]{0.45\textwidth}
        \begin{tikzpicture}
        	\node[inner sep=0pt] at (0,0)
        	{\includegraphics{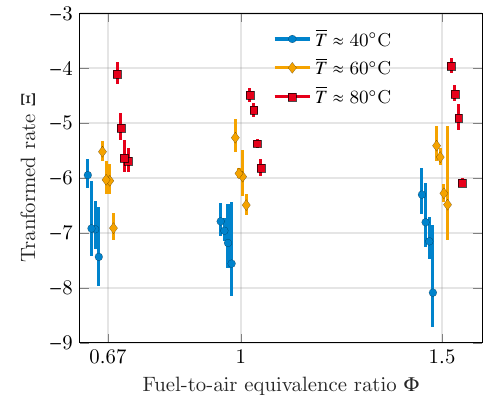}};
        \end{tikzpicture}
        \caption{Effect of $\Phi$ on the transformed rate $\Xi$.\\
        ~~\\
        ~~}
        \label{fig:main_effects_oxs}
    \end{minipage}    
    \hspace{0.8cm}
    \begin{minipage}[b]{0.45\textwidth}
        \begin{tikzpicture}
        	\node[inner sep=0pt] at (0,0)
        	{\includegraphics{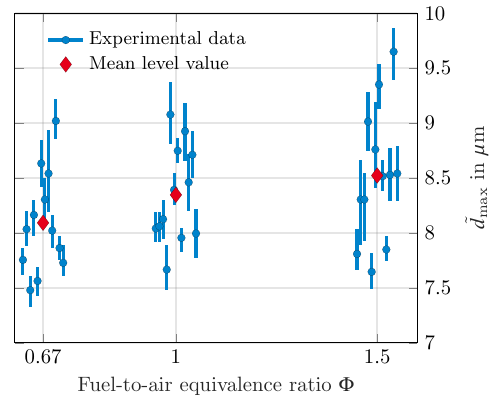}};
        \end{tikzpicture}
        \caption{Maximum median particle diameter $\Tilde{d}_{\mathrm{max}}$ as a function of $\Phi$. The error bars indicate the $\tilde{d}_{95,\mathrm{boot}}$ at $t(\Tilde{d}=\Tilde{d}_{\mathrm{max}}$.}
        \label{fig:oxs_diameter}
    \end{minipage}
\end{figure}
\newpage
\section{Speciation of iron oxalate complexes}
\label{sec:speciation_of_iron_oxalate_complexes}
\begin{figure}[!htbp]
    \begin{minipage}[t]{0.49\textwidth}
    \begin{tikzpicture}
        \node[inner sep=0pt] at (0,0)
         {\includegraphics[trim=0 0 0 0,clip,width=1\textwidth]{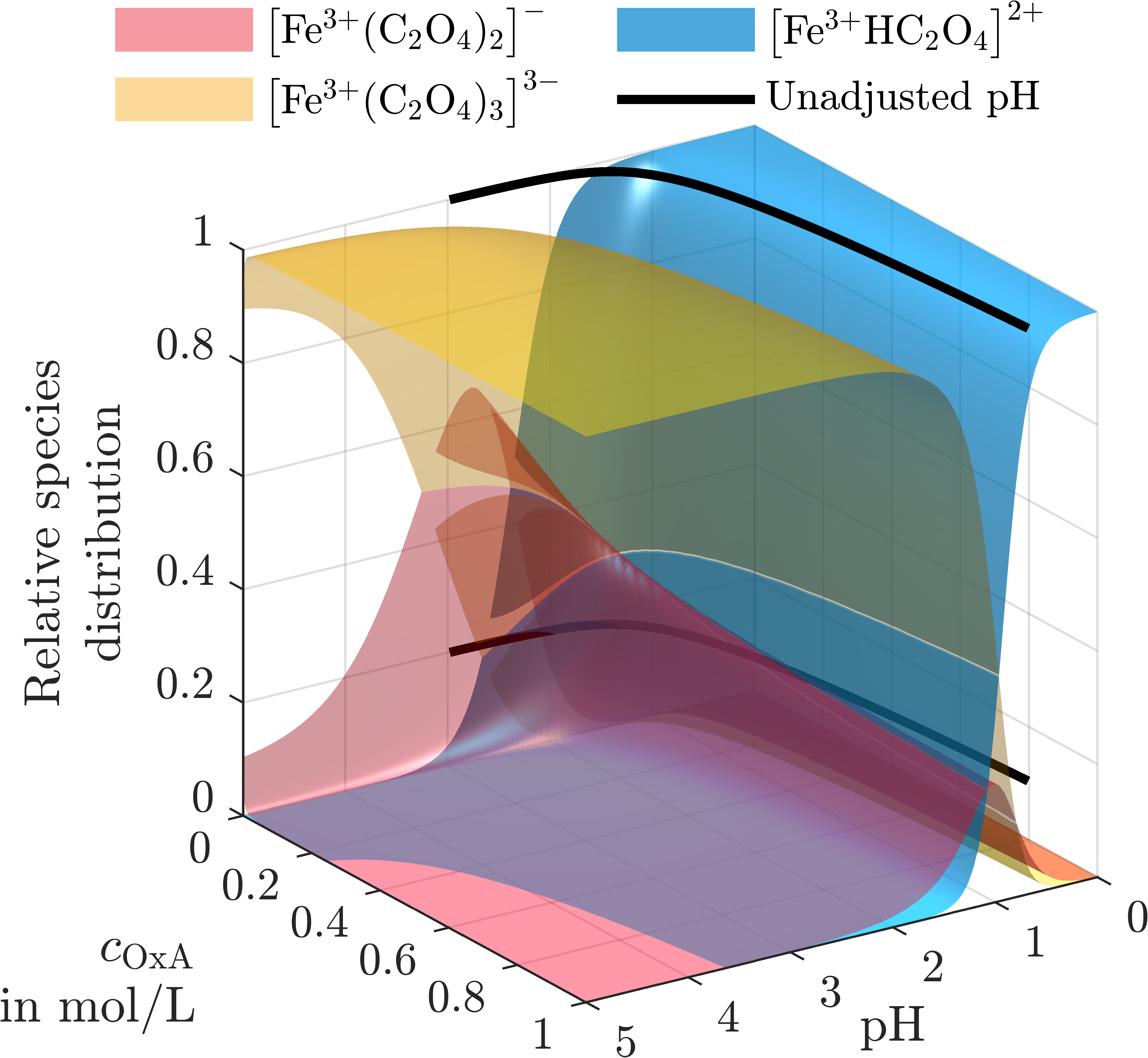}};
        \node[align=center,fill=white,draw=none] at (-3.6,2.6) {(a)};
    \end{tikzpicture}
    \end{minipage}
    \hfill
    \begin{minipage}[t]{0.49\textwidth}
    \begin{tikzpicture}
        \node[inner sep=0pt] at (0,0)
        {\includegraphics[trim=0 0 0 0,clip,width=1\textwidth]{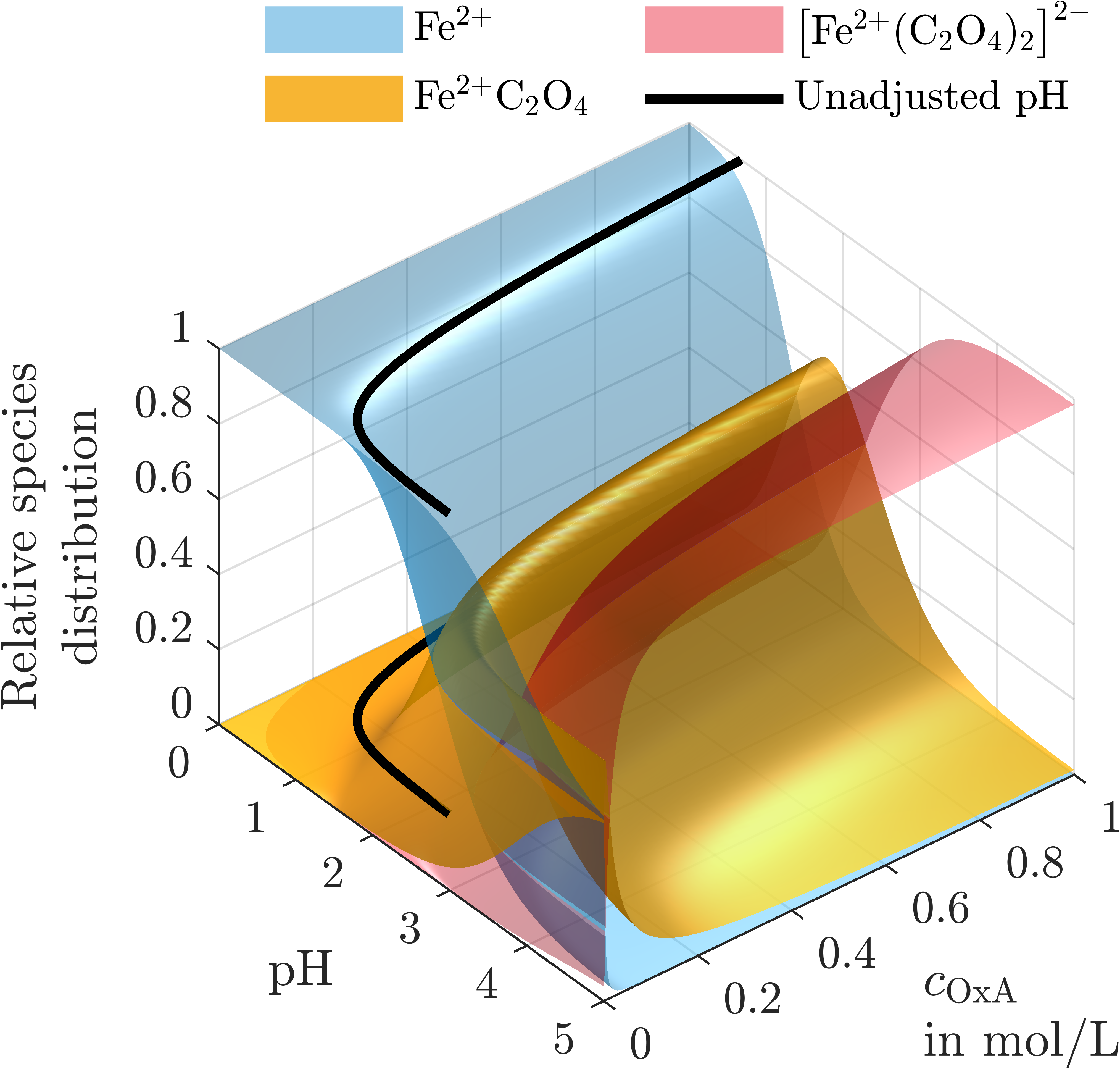}};
        \node[align=center,fill=none,draw=none] at (-4.0,2.6) {(b)};
    \end{tikzpicture}
	\end{minipage}
    \caption{Relative species distribution of (a) ferric and (b) ferrous complexes as a function of pH and $c_{\mathrm{OxA}}$. In (a), non\textendash complexed $\mathrm{Fe}^{3+}$ were omitted due to their negligible contribution. The black lines represent the unadjusted pH at that OxA concentration.}
\label{fig:speciation_all_fe2_fe3}
\end{figure}

\begin{figure}[ht!]
\centering
        \begin{tikzpicture}
            \node[inner sep=0pt] at (0,5)
            {\setlength\fwidth{0.52\textwidth}
            \setlength\fheight{0.32\textwidth}
            \input{Fig_E21a_OxA_Fe3_complexes_P1}};
            \node[align=center,fill=none,draw=none] at (-3.85,6.6) {(a)};
        \end{tikzpicture}
        ~
        \begin{tikzpicture}
            \node[inner sep=0pt] at (0,0)
            {\setlength\fwidth{0.55\textwidth}
            \setlength\fheight{0.4\textwidth}
            \input{Fig_E21b_OxA_Fe2_complexes_P1}};
            \node[align=center,fill=none,draw=none] at (-4.3,1.9) {(b)};
        \end{tikzpicture}
    \caption{Relative distribution of (a) ferric and (b) ferrous species as a function of pH, calculated using the equilibrium constant values from \citet{PANIAS1996B} and \citet{POZDNYAKOV2008}. The pH at $c_{\mathrm{OxA}}=0.19\,$M was calculated assuming the dissociation constants $\mathrm{pKa_1} = 1.25$ and $\mathrm{pKa_2} = 4.21$ at 25$^\circ$C~\citep{PANIAS1996A}. The data points represent the measured fraction of total dissolved iron as a function of pH at $c_{\mathrm{OxA}}=0.19\,$M, obtained from \citet{LEE2007}. Reproduced with permission from Elsevier.}
    \label{fig:speciation_Lee_concentration}
\end{figure}

\newpage
\section{Supplementary experimental data}
\label{sec:supplementary_exp_data}

\begin{figure*}[ht!]
\centering
    \begin{minipage}[b]{0.4\textwidth}
    	\begin{tikzpicture}
    		\node[inner sep=0pt] at (0,0)
    		{\includegraphics{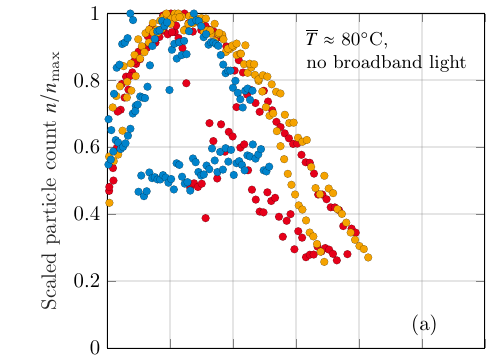}};
    	\end{tikzpicture}
    \end{minipage}  
    \hspace{0.5cm}
    \begin{minipage}[b]{0.4\textwidth}
    	\begin{tikzpicture}
    		\node[inner sep=0pt] at (0,0)
    		{\includegraphics{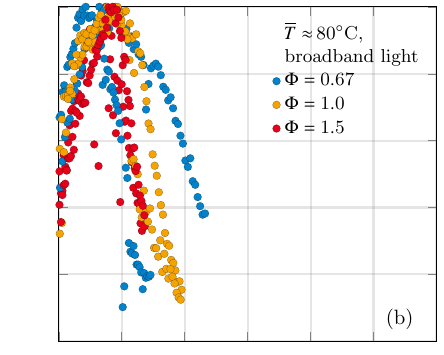}};
    	\end{tikzpicture}
    \end{minipage}
    ~\\
    \begin{minipage}[b]{0.4\textwidth}
    	\begin{tikzpicture}
    		\node[inner sep=0pt] at (0,0)
    		{\includegraphics{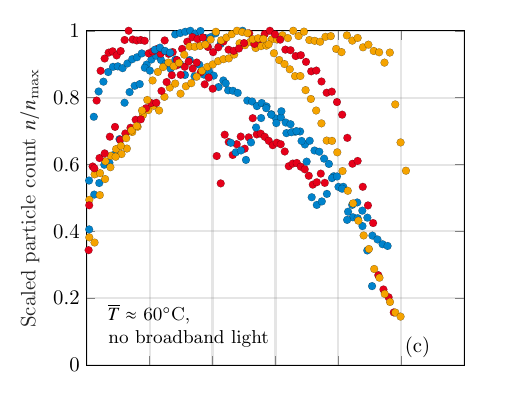}};
    	\end{tikzpicture}
    \end{minipage}
    \begin{minipage}[b]{0.4\textwidth}
    	\begin{tikzpicture}
    		\node[inner sep=0pt] at (0,0)
    		{\includegraphics{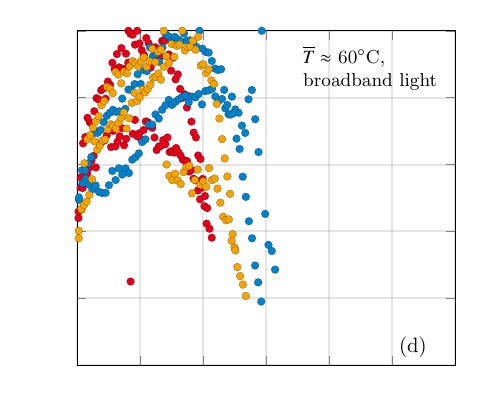}};
    	\end{tikzpicture}
    \end{minipage}
    ~\\
    \begin{minipage}[b]{0.4\textwidth}
    	\begin{tikzpicture}
    		\node[inner sep=0pt] at (0,0)
    		{\includegraphics{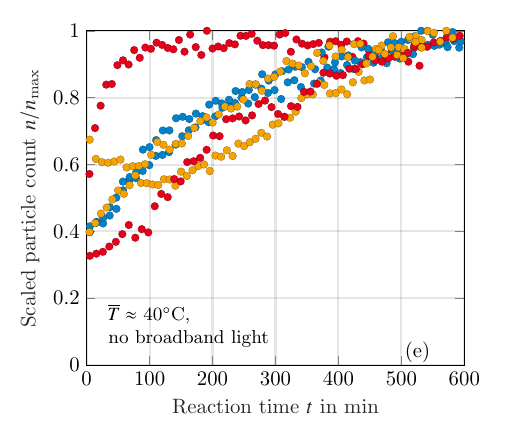}};
    	\end{tikzpicture}
    \end{minipage}
    \begin{minipage}[b]{0.4\textwidth}
    	\begin{tikzpicture}
    		\node[inner sep=0pt] at (0,0)
    		{\includegraphics{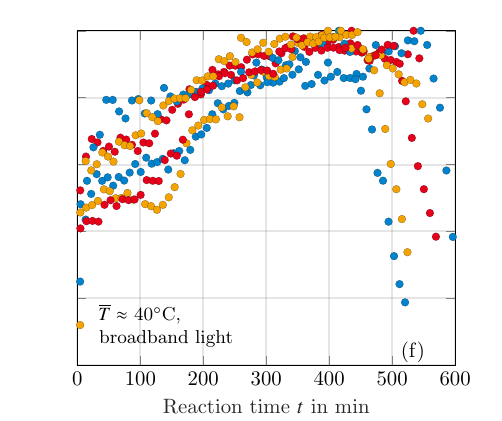}};
    	\end{tikzpicture}
    \end{minipage}
    \caption{Number of detected particles $n(t)/n_{\mathrm{max}}$ at $\overline{T}=80^\circ$C (a,b), 60$^\circ$C (c,d) and 40$^\circ$C (e,f), both without (a,c,e) and with broadband (b,d,f) light irradiation for varying $\Phi$ (0.67, 1.0, 1.5), each measured two times.}
    \label{fig:raw_data_all_n}
\end{figure*}

\begin{figure}[ht!]
    \centering
    \begin{tikzpicture}
       	\node[inner sep=0pt] at (0,0)
        	{\includegraphics{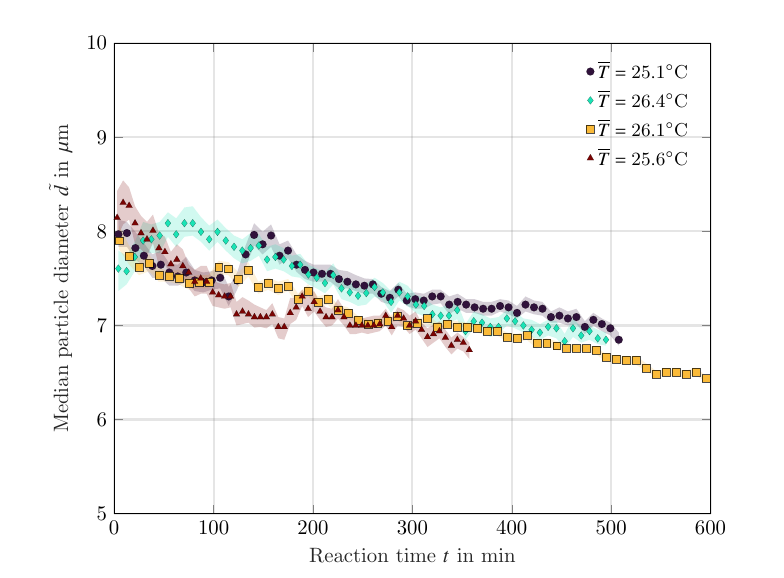}};
    \end{tikzpicture}
    \caption{Median particle diameter $\Tilde{d}(t)$ at room temperature without broadband light irradiation for $\Phi=0.67$.}
    \label{fig:fragmentation_diameter_evolution}
\end{figure}

\begin{figure}[ht!]
    \centering
    \begin{tikzpicture}
       	\node[inner sep=0pt] at (0,0)
        	{\includegraphics{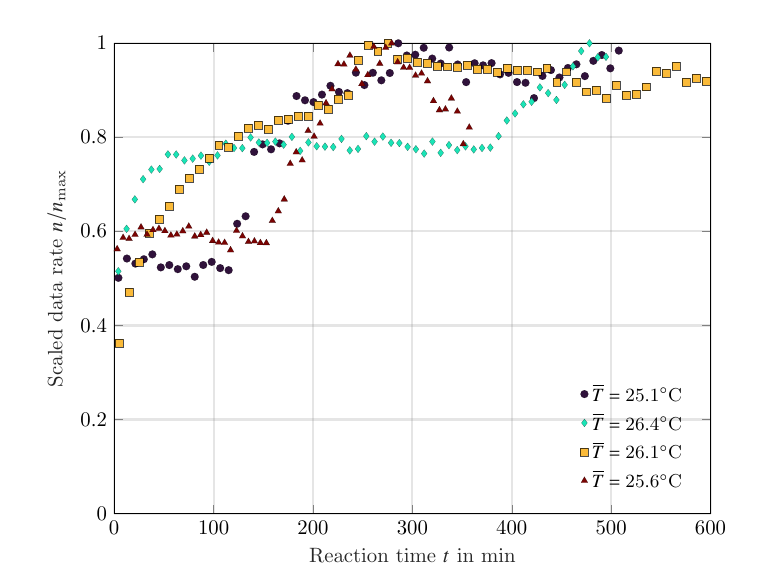}};
    \end{tikzpicture}
    \caption{Number of detected particles $n(t)/n_{\mathrm{max}}$ at room temperature without broadband light irradiation for $\Phi=0.67$.}
    \label{fig:fragmentation_n_evolution}
\end{figure}

\end{document}